\documentclass[preprint,floats,tightenlines,11pt,eqsecnum,aps]{revtex4}
\usepackage{amsmath}
\usepackage{graphicx}
\begin{document}
\def\appls{\hbox{$<$\kern-.75em\lower 1.00ex\hbox{$\sim$}}}
\title{EVIDENCE FOR EVOLUTION FROM PURE STATES TO MIXED STATES\\
IN PION CREATION PROCESS $\pi^- p\to \pi^-\pi^+ n$ ON POLARIZED TARGET\\
 AND ITS PHYSICAL INTERPRETATION.}

\author{Miloslav Svec\footnote{electronic address: svec@hep.physics.mcgill.ca}}
\affiliation{Physics Department, Dawson College, Montreal, Quebec, Canada H3Z 1A4}
\date{August 28, 2007}

\begin{abstract}

In 1982 Hawking suggested that quantum fluctuations of the space-time metric will induce a non-unitary evolution from pure initial states $\rho_i$ to mixed final states $\rho_f$ in particle interactions at any energy. This hypothesis can be tested using existing CERN data on $\pi^- p \to \pi^- \pi^+ n$ on polarized target at 17.2 GeV/c. The purity of the final state $\rho_f$ is controlled by the purity of the recoil nucleon polarization. We develop spin formalism to calculate the expressions for recoil nucleon polarization for two specific measured initial pure states. Imposing the condition of purity on the recoil nucleon polarization we obtain conditions on the amplitudes which are violated by model independent amplitude analyses of the CERN data on polarized target at large momentum transfers. We conclude that pure states can evolve into mixed states in $\pi^- p \to \pi^- \pi^+ n$. In quantum theory such non-unitary evolution occurs in open quantum systems $S$ interacting with a quantum environment $E$. The reduced density state is mixed and is given by Kraus representation
$\rho_f(S)=Tr_E(\rho_f(S,E))=\sum \limits_\ell S_\ell \rho_i(S)S^+_\ell$. We show that Kraus representation leaves invariant the formalism used in data analyses provided the co-evolution with the environment conserves $P$-parity and quantum numbers of the environment. The measured density matrix elements are redefined to be environment-averaged density matrix elements. Measured elements that depend explicitely on the environment are predicted to violate certain phase relations. Measured elements that do not depend explicitely on the environment form a decoherence free subspace and satisfy the phase relations. This prediction is in excellent agreement with the CERN data and validates the view of pion creation processes as open quantum systems interacting with a quantum environment. The interaction can be thought of as a scattering of hadrons carrying energy-momentum and spin with particles of the environment carrying quantum entanglement. There is no exchange of energy-momentun between the hadrons and the environment in this non-dissipative, entanglement changing interaction. The observed process is time-irreversible and violates $CPT$ symmetry. Following the proposal by Hawking, we suggest the origin of the environment is in quantum gravity rendered observable by the pion creation process. 

\end{abstract}

\maketitle

\tableofcontents

\newpage

\section{Introduction: Spin physics and the unitary $S$-matrix.}

The concept of scattering matrix was introduced by Wheeler in 1937~\cite{wheeler37}. In 1943 Heisenberg introduced the concept of $S$-matrix to describe observables in interactions of elementary particles~\cite{heisenberg43}. Heisenberg understood the $S$-matrix as a matrix of transition probability amplitudes between the initial and final states of the interactions. Just like the transitions of electrons between energy levels in atoms are not observable processes, the process of interactions of elementary paricles is not accessible to direct observation. Heisenberg believed that these classically unobservable processes are deeply related to our understanding of spacetime and will be eventually described by some sort of non-linear field theory that may encompass the irreversibility of quantum mechanics measurements~\cite{vialtzev65}. In the absence of such a theory he proposed as a tentative solution the idea of asymptotic $S$-matrix defined in terms of in- and out- states in an analogy with potential scattering in quantum mechanics~\cite{vialtzev65}. The unitarity of the $S$-matrix was demonstrated by Pauli in 1946~\cite{pauli46}. Although Heisenberg believed that asymptotic $S$-matrix is a temporary tool, the success of Quantum Electrodynamics developed by Feynman, Schwinger and Tomonaga in 1940's provided evidence for its utility and meaningful physical interpretation. All Quantum Field Theories that followed were modeled on QED. To relate the theory to measurable $S$-matrix amplitudes, a reduction formalism is first used to express the $S$-matrix elements in terms of product of vacuum expectation values of fully interactng fields~\cite{kaku93,itzykson05}. Next a unitary time evolution operator $U(t)$ is found that takes the fully interacting fields to the free, asymptotic fields
\begin{equation}
U(t)=U(t,- \infty) = T exp \bigl ( - \int_{-\infty}^t dt H_I(t)\bigr )=
T exp \bigl ( - \int_{-\infty}^t dt \int d^3x ~{\cal {H}}_I (\vec {x},t) \bigr )
\end{equation}
\noindent
where $T$ is time ordering operator and $H_I(t)$ is the interaction Hamiltonian defined only with free, asymptotic fields. The $S$-matrix elements are then expressed in terms of vacuum expectation values of free fields which allows their calculation. The asymptotic $S$-matrix is defined as the limit
\begin{equation}
S= \lim \limits_{t \to + \infty} U(t, -\infty)=U(+ \infty, - \infty)
\end{equation}
\noindent
The evolution operator $U(t)$, and thus the unitary $S$-matrix, evolves pure isolated initial state into a pure isolated final state. Since the unitary evolution $U(t)$ is time reversible, the interactions of particles must be time reversible isolated processes.\\

In 1980 Wald showed rigorously that any scattering process of particles that evolves pure initial state into a mixed final state violates $CPT$ symmetry and is time 
irreversible~\cite{wald80}. He suggested that such processes will occur in curved space-time, and that quantum gravity violates $CPT$ symmetry and time-reversal invariance. In 1982 Hawking showed that in the presence of a black hole (macroscopic or microscopic) a pure initial state of interacting particles will evolve into a mixed final state as some of the quantum states produced in the particle interaction will fall behind the horizon and become inaccessible to quantum measurements by the observer~\cite{hawking82}. He also suggested that quantum fluctuations of the space-time metric will have the same effect on interacting particles and induce their non-unitary evolution - at any energy~\cite{hawking82,hawking84}.
Hawking questioned a universal validity of the unitary time evolution (1.1) in the presence of metric fluctuations and suggested that initial and final state density matrices $\rho_{in}$ and $\rho_{out}$ are connected by a linear but non-unitary evolution operator. Such mappings in fact exist and describe non-unitary evolution $\rho_{out}(S)={\cal E}(\rho_{in}(S))$ of an open quantum system or process $S$ interacting with an environment 
$E$~\cite{kraus83,blum96,nielsen00,breuer02}.\\\\

Hawking's ideas inspired suggestions to test them experimentally. In 1984, Ellis, Hagelin,
Nanopoulos and Srednicki proposed that quantum fluctuations of the metric form an environment with which interacting as well as free particles interact as open quantum 
systems~\cite{ellis84}. Such interactions would lead to an observable loss of coherence and $CPT$ violations in $K^0 \overline{K}^0$ systems which maintain coherence over macroscopic distances. They supplemented Limblad time evolution equation for density matrix in dissipative open quantum systems with $CPT$ violating terms to model interaction with quantum fluctuations and estimated the magnitude of their effects. Over the years and up to very recently  other suggestions to test Hawking's ideas have been put 
forward~\cite{huet95,ellis96,gerber98,gerber04,mavromatos06,bernabeu06a,bernabeu06b,koranga06}. During the recent years experiments with neutral kaons have yielded remarkably sensitive results on violations of $CPT$ symmetry and time reversal invariance, coherence of wave functions  and entanglement of kaon pairs~\cite{fidecaro06}. So far these experiments did not provide a conclusive confirmation of a non-unitary evolution of free kaon systems, and thus possible evidence for quantum gravity effects.\\

In this work we return to the original idea of Hawking that pure initial state of interacting particles will evolve into a mixed final state as the result of interacctions with quantum fluctuaions of the metric during the process of particle interaction. To test unitarity in particle scattering requires to examine a variety of exclusive hadron processes with known initial and final state density matrices measured in spin physics experiments using polarized targets or polarized beams. Spin physics experiments make no use of the unitarity assuptions (1.1) and (1.2) and can thus inform us about their validity.\\

Spin physics as a research field was initiated in 1946 and 1948 by Julian Schwinger. Schwinger wanted to know if nuclear forces are indeed spin independent as was generally assumed at that time. He designed experimental methods to create polarized neutron beams and to measure their polarization~\cite{schwinger46,schwinger48}. In 1949 Wolfenstein introduced the concept of spin observables measurable in Schwinger's experiments with nucleons~\cite{wolfenstein49}. In 1957 Chamberlain and collaborators at Berkeley reported the first evidence for spin dependence of proton-proton and proton-neutron scattering at 320
MeV ~\cite{chamberlain57}. An important element of their experiments was the measurement of recoil nucleon polarization in a secondary rescattering, a method first proposed by Schwinger.\\ 

In 1961 Bethe and Schumacher introduced the concept of amplitude analysis - the construction of scattering matrix from the measurements of a complete set of spin 
observables~\cite{schumacher61}. In 1972 van Rossum, an earlier associate of Chamberlain at Berkeley, and his spin physics group at CEN-Saclay reported the first measurements of recoil nucleon polarization in $\pi^+ p$ elastic scattering at 6 GeV/c at CERN~\cite{delesquen72}. These measurements closed the set of spin observables and enabled them to perform the first amplitude analysis of a hadronic reaction. The measured pion-nucleon amplitudes at 6 GeV/c~\cite{cozzika72} invalidated all Regge models, and emphasized the importance of experimental measurements of scattering amplitudes in experiments using polarized targets.
In 1978 Lutz and Rybicki extended the concept of amplitude analysis to pion production processes $\pi N \to \pi^- \pi^+ N$ and showed that a nearly complete amplitude analysis is possible using only measurements on transversely polarized target~\cite{lutz78}. Their results were later generalized to $\pi^- p \to \pi^0 \pi^0 n$ and $\pi^- p \to \pi \eta n$ processes~\cite{svec97d,svec97e}. Amplitude analysis formalism was also extended to inclusive measurements~\cite{doncel72,goldstein76}. A comprehensive introduction to spin in particle physics and modern spin physics technology is given in Ref.~\cite{bourrely80} and in the recent book by Leader~\cite{leader01}.\\

The result of any possible experiment of a considered reaction $|i> \to |f>$ involving measurements with spin of particles is the measurement of its final state density matrix $\rho_f$. Given the initial state density matrix $\rho_i$, the final state $\rho_f$ is given by a quantum evolution of the initial state $\rho_i$ determined by the $S$-matrix 
\begin{equation}
\rho_f=S \rho_i S^+
\end{equation}
\noindent
The equation (1.3) allows us to express $\rho_f$ in terms of the full set of transition amplitudes describing the reaction. At first we are interested in two processes $\pi^- p \to \pi^- p$ and $\pi^- p \to \pi^- \pi^+ n$. As we shall see later, the purity of the initial 
state $\rho_i(\pi^- p)$ is controlled by target polarization vector $\vec {P}$ while the purity of the final states $\rho_f(\pi^- p)$ and $\rho_f(\pi^- \pi^+ n)$ is controlled by recoil nucleon polarization vectors $\vec {Q}(\vec {P})$ and $\vec {Q}(\Omega, \vec {P})$, respectively. Here $\Omega = (\theta, \phi)$ specifies the direction of $\pi^-$ in the center of mass systems of the two pions. The initial state is pure if and only if $|\vec {P}|^2=1$.
Similarly, the final states are pure if and only if the recoil polarizations have a unit magnitude $|\vec {Q}(\vec {P})|^2=1$ and  $|\vec {Q}(\Omega, \vec {P})|^2=1$~\cite{nielsen00}.\\

Using known expressions for recoil nucleon polarization in $\pi^- p \to \pi^- p$~\cite{perl74} we find that pure states evolve always into pure states. The same conclusion holds true for a class of meson baryon two-body reactions with the same spin structure. Such is not the case for $\pi^- p \to \pi^- \pi^+ n$.\\

Following the initial work of Lutz and Rybicki~\cite{lutz78}, we develop a spin formalism in
$\pi^- p \to \pi^- \pi^+ n$ to express the final state density matrix and recoil nucleon polarization in terms nucleon transversity amplitudes with a definite dipion spin, helicity and naturality. We then calculate exact expressions for $|\vec {Q}(\Omega, \vec {P})|^2$ for two special pure initial states with transverse polarizations $P_y=\pm 1$ and impose the condition that the corresponding final states is pure. For each $P_y= \pm 1$ we obtain an elegant relationship between moduli and certain relative phases of all amplitudes that must be both satisfied by the unitary $S$-matrix. These unitarity conditions are satisfied for dimeson masses where only $S$- and $P$-wave contribute. At higher masses the unitarity conditions require that either (A) relative phases between natural (unnatural) exchange amplitudes of the same nucleon transversity are all zero or (B) natural (unnatural) exchange amplitudes with dimeson spins larger than 1 all vanish. These conditions must hold true at all energies $s$, dimeson masses $m$ and momentum transfers $t$. The unitarity conditions (A) and (B) also apply to a class of similar processes with dimeson final states that have the same spin structure, such as $\pi N \to \pi^0 \pi^0 N$, $KN \to K \pi N$, $\pi N \to K \pi \Lambda^0$ etc..\\

CERN measurements of $\pi^- p \to \pi^- \pi^+ n$ on polarized target at 17.2 GeV/c find non-zero unnatural exchange amplitudes in $S$-, $P$- and $D$-waves and non-zero natural exchange amplitudes in $P$- and $D$-waves at small~\cite{becker79b,chabaud83} as well as at large~\cite{rybicki85} momentum transfers $t$. These measurements also find non-zero phases between all unnatural exchange amplitudes and between the $P$-wave and $D$-wave natural exchange amplitudes, at small and large $t$. Both unitarity conditions (A) and (B) are thus violated by the CERN measurements. We conclude that in $\pi^- p \to \pi^- \pi^+ n$ and in similar pion creation processes pure states can evolve into mixed states in a non-unitary evolution of initial states $\rho_i$ into final states $\rho_f$.\\

In quantum theory non-unitary evolutions occur in open quantum systems interacting with an environment~\cite{kraus83,blum96,nielsen00,breuer02}. The co-evolution of the system $S$ with a quantum environment $E$ is a quantum operation $\rho_f(S,E)=U \rho_i(S,E)U^+$ with a unitary evolution operator $U$. The initial state of the combined system is assumed to be separable state $\rho_i(S,E)=\rho_i(S)\otimes \rho_i(E)$ but the the final state $\rho_f(S,E)$ is an entangled state of $S$ and $E$. The measured state $\rho_f(S)$ is a reduced state given by Kraus represetation
\begin{equation}
\rho_f(S)=Tr_{E}(\rho_f(S,E)=\sum \limits_{\ell} S_\ell \rho_i(S) S^+_\ell
\end{equation}
where $S_\ell=<e_\ell|U|e_0>$ are operators acting on the state space of the system $S$, $|e_\ell>$ are quantum states of the environment, and where we assumed that the initial state is a pure state $\rho_i(E)=|e_0><e_0|$. Kraus representation describes a non-unitary evolution from $\rho_i(S)$ to $\rho_f(S)$, and the state $\rho_f(S)$ is a mixed state.\\

At first sight there appears to be a contradiction. Starting with the unitary evolution (1.3), a spin formalism is developed to analyze the measured data. The data analysis leads to conclusion that the evolution is non-unitary and described by Kraus representation (1.4). We show that Kraus representation leaves the form of all equations invariant provided that the co-evolution $U$ conserves $P$-parity and quantum numbers of the environment. However, the measured density matrix elements and the measured moduli and phases are environment-averaged values of co-evolution density matrix elements and co-evolution amplitudes that explicitely depend on interacting degrees of the environment. The conclusion, that the measured density matrix elements are averaged values predicts a violation of certain phase relations. We show that such phase relations are violated by the CERN data on $\pi^- p \to \pi^- \pi^+ n$.\\

The CERN data validate the Kraus representation and the new view of pion creation processes as open quantum systems interacting with a quantum environment. We can think of the co-evolution process as a scattering of initial hadron states that carry energy-momentum and spin with particles of the environment that carry quantum entanglement. There is no exchange of energy-momentum between hadrons and the environment in this non-dissipative process. Since environment states do not have well defined antiparticle states and the interaction is non-local, the coevolution process violates $CPT$ symmetry. Since the measured process is non-unitary, it is time-irreversible and violates $CPT$ symmetry. Following the proposals by Wald~\cite{wald80} and Hawking~\cite{hawking82,hawking84}, we suggest that the observed quantum environment originates in quantum gravity. In this view, pion creation processes act as non-classical instruments that make observations of quantum gravity possible.\\ 

The paper is organized as follows. In Section II. we present the form of the final states and conditions for its purity in $\pi^- p \to \pi^- \pi^+ n$ and develop the necessary spin formalism that will be used in this work and in sequel papers. In Section III. we show that pure states evolve into pure states in $\pi N \to \pi N$ and in similar processes. In Section IV. we derive conditions that must be satisfied by the amplitudes in $\pi^- p \to \pi^- \pi^+ n$ in order for two specific pure initial states to evolve into pure final states. In Section V. we show that these conditions are violated by CERN data on $\pi^- p \to \pi^- \pi^+ n$ on polarized target and conclude that in $\pi^- p \to \pi^- \pi^+ n$ pure states can evolve into mixed states. In Section VI. we comment on a formal connection between a change from unitary to non-unitary evolution and a change in metric. In Section VII. we briefly review the basic  concepts of evolution of open quantum systems interacting with environment and Kraus representation. In Section VIII. we show that Kraus representation leaves invariant the spin formalism developed in Section II. provided that the co-evolution with environment conserves $P$-parity and quantum numbers of the environment states. The prediction that the measured density matrix elements and bilinear terms of amplitudes are environment-averaged values is validated in Section IX by the CERN data, validating the view of pion creation processes as open quantum systems interacting with environment. In Section X. we introduce a concept of decoherence free subspace that is central in determination of the quantum states of the environment, briefly discussed in Section XI. The time-irreversibility and $CPT$ violation in pion creation processes is discussed in Section XII. The paper closes with conclusions in Section XIII. Appendix A presents the Lutz-Rybicki tables of relations for density matrix elements in terms of amplitudes on which the present work is based.

\section{Final state density matrix in $\pi^- p \to \pi^- \pi^+ n$ and recoil nucleon\\ polarization.}

\subsection{From $S$-matrix to transition matrix $T$.}

In the laboratory system of the reaction $\pi^- p \to \pi^- \pi^+ n$ the $+z$ axis has the direction opposite to incident pion beam. The $+y$ axis is perpendicular to the scattering plane and has direction of $\vec {p}_{\pi^-} \times \vec {p}_{\pi^- \pi^+}$. The angular distribution of produced dipion system is described by the direction of $\pi^-$ in the dipion center-of-mass system and its solid angle $\Omega = \theta, \phi$. The target nucleon and recoil nucleon helicities $\nu$ and $\chi$ are defined in $s$-channel helicity system. The dipion helicity $\lambda$ will be defined in the $t$-channel helicity system~\cite{lutz78,becker79a,lesquen85,lesquen89}.\\

The pion beam and nucleon target are prepared in a separable state $\rho_i(\pi^- p) = \rho_i(\pi^-) \otimes \rho_i(p)$ where $\rho_i(\pi^-) = |p_{\pi}0 > <p_{\pi}0| \otimes |1,-1><1,-1|$ and
\begin{equation}
\rho_i(p)=(\sum \limits_{\nu \nu^{'}} (\rho^{1 \over{2}}_p (\vec{P}))_{\nu \nu^{'}} |p_p \nu> <p_p \nu^{'}|) \otimes |{1 \over{2}},+{1 \over{2}}><{1 \over{2}},+{1 \over{2}}|
\end{equation}
$|p_{\pi} 0>$ and $|p_p \nu>$ are pion and proton helicity state vectors, respectively, and $|I,I_3><I,I_3|$ are their isospin states; 0 stands for pion helicity. $\rho^{1 \over{2}}_p (\vec{P})$ is the target spin density matrix. In the following we suppress the momentum and isospin labels in the initial and final helicity states $|0 \nu>$ and $|\theta \phi, \chi>$. The final state vectors form a basis of orthonormal state vectors in the recoil nucleon spin space $<\theta \phi, \chi| \theta \phi, \chi^{'}> = \delta_{\chi \chi^{'}}$. The final state density matrix $\rho_f(\pi^- \pi^+ n)$ 
\begin{equation}
\rho_f(\theta \phi, \vec{P})=\sum \limits_{\chi \chi^{'}} (\rho_f(\theta \phi, \vec{P}))^{{1 \over{2}}{1 \over{2}}}_{\chi \chi{'}} |\theta \phi, \chi><\theta \phi, \chi^{'}|
\end{equation}
\noindent
has matrix elements given by evolution equation $\rho_f=S \rho_i S^+$
\begin{equation}
(\rho_f(\theta \phi, \vec{P}))^{{1 \over{2}}{1 \over{2}}}_{\chi \chi'}=\sum \limits_{\nu \nu'} <\theta \phi, \chi|S|0 \nu>(\rho^{1 \over{2}}_p (\vec{P}))_{\nu \nu'} 
<0 \nu'|S^+|\theta \phi, \chi'>
\end{equation}  
\noindent
We will use an abbreviated notation for amplitudes $S_{\chi, 0 \nu}=<\theta \phi, \chi|S|0 \nu>$ and suppress their dependence on energy $s$, momentum transer $t$, dipion mass $m$ and the angles $\theta \phi$. We also drop the superscripts ${1 \over{2}}{1 \over{2}}$.\\

Transition $T$-matrix is defined by $S_{fi}=I_{fi}+i(2 \pi)^4 \delta^4 (P_f-P_i) T_{fi}$. With
amplitudes $S_{\chi, 0 \nu}=iT_{\chi, 0 \nu}(2 \pi)^4 \delta^4 (P_f-P_i)$ we get
\begin{equation}
\rho_f(\theta \phi, \vec{P})_{\chi \chi'}=\rho^{'}_f(\theta \phi, \vec{P})_{\chi \chi'}(VT) (2\pi)^4 \delta^4(P_f-P_i)
\end{equation}
\noindent
where $\rho^{'}_f(\theta \phi, \vec{P})$ is expressed in terms of transition amplitudes $T_{\chi, 0 \nu}$ and where we have used the conventional approach to deal with a square of $\delta$-functions~\cite{perl74} with $V$ and $T$ being total volume and time confining the interactions to be taken in the limit $V,T \to \infty$. According to the Born rule, the probability of $\pi^- p(\nu) \to \pi^-(p_1) \pi^+(p_2) n(p_3,\chi)$ is given by
\begin{equation}
dP_{\chi,0\nu}=|S_{\chi,0\phi}|^2 \prod \limits^3_{n=1} {d^3 \vec{p}_n \over{(2\pi)^3 E_n}}=
|T_{\chi,0\phi}|^2 d\Phi_3(P_i,p_1,p_2,p_3)(VT)
\end{equation}
\noindent
where the Lorentz invariant phase space $d\Phi_3=q(m^2)G(s)dmdtd\Omega$ with $q(m^2)$ and $G(s)$ the pion momentum in dipion center-of-mass system and $G(s)$ the energy dependent part of the phase space~\cite{svec97a}. The probability per unit volume, unit time and per target particle is $d\sigma_{\chi,0\nu}=dP_{\chi,0\nu}/(VTFlux(s))$ and the differential cross-section reads
\begin{equation}
{d\sigma_{\chi,0\nu} \over {dtdmd\Omega}}={q(m^2)G(s) \over {Flux(s)}}|T_{\chi,0\phi}|^2
\end{equation}
\noindent
Applying formally the same procedure to every bilinear term $S_{\chi^{'},0\nu^{'}}S_{\chi,0\nu}^*$ of $\rho_f(\Omega ,\vec{P})$ we can define a differential cross-section matrix 
\begin{equation}
{d\sigma \over {dtdmd\Omega}}={q(m^2)G(s) \over {Flux(s)}}\rho^{'}_f(\Omega ,\vec{P}) \equiv 
\rho_f(\Omega ,\vec{P})
\end{equation}
\noindent
where we have absorbed $\sqrt {q(m^2)G(s)/Flux(s)}$ into transition amplitudes and redefined $\rho_f(\Omega ,\vec{P})$. It can be written in a matrix form
\begin{equation}
 \rho_f = T \rho_i  T^+
\end{equation}
\noindent
where $T$ is the matrix of transition amplitudes. The transition matrix $T$ is non-unitary and non-hermitian but it still evolves pure initial states into pure final states on account of the central assumption (1.1) and unitary $S$-matrix (1.2).

\subsection{Conditions for purity of initial and final states.}

The target nucleon spin density matrix has the form~\cite{bourrely80,leader01}
\begin{equation}
\rho_p (\vec {P}) = {1 \over{2}} (1+\vec{P} \vec{\sigma})
\end{equation}
\noindent
where $\vec{P}$ is the target polarization vector and $\vec{\sigma} = (\sigma_x, \sigma_y, \sigma_z)$ are Pauli matrices. The target is in a pure state if and only if $|\vec{P}|^2=1$; otherwise it is in a mixed state~\cite{nielsen00,leader01}. When $\vec{P}=0$ the target is maximally mixed state with equal probability of the target nucleon spins "up" or "down" relative to the scattering plane, and the spin ensemble is isotropic. When $\vec{P} = (0, \pm P, 0)$, the target is transversely polarized with spins "up" for $+P$ or "down" for $-P$, respectively. For $P= \pm 1$ the transversely polarized target is in a pure state 
$\rho_{p,u}={1 \over{2}}(1+\sigma_y)$ or $\rho_{p,d}={1 \over{2}} (1-\sigma_y)$. In modern polarized targets the initial density matrix $\rho_i(\vec {P})$ can be varied by external magnetic fields to rotate the polarization vector $\vec{P}$ into any desired direction.\\

To discuss the purity of the final state we note a useful result from quantum state tomography~\cite{nielsen00}. Arbitrary density matrix $\rho$ of $n$ qubits can be expanded in a form 
\begin{equation}
\rho = \sum \limits_{\vec{v}} ({1 \over{2^n}}) Tr(\sigma_{v_1} \otimes \sigma_{v_2} \otimes ... \otimes \sigma_{v_n} \rho) \sigma_{v_1} \otimes \sigma_{v_2} \otimes ... \otimes \sigma_{v_n}
\end{equation}
\noindent
where the sum is over the vectors $\vec{v} = (v_1, v_2, ..., v_n)$ with entries chosen from the set $\sigma^j, j=0,1,2,3$ of Pauli matrices and $\sigma^0 = 1$. This result can be generalized to any non-qubit systems~\cite{nielsen00}. The traces in (2.10) represent average values of spin correlations. The determination of these averages requires repeated measurements forming a large ensemble of events. The expansion (2.10) shows that the concept of density matrix in Quantum Theory is inherently a statistical concept. The final density matrix $\rho_f(\theta \phi, \vec{P})$ is a single qubit density matrix corresponding to spin ${1 \over{2}}$ of the recoil nucleon. It can be written in the form (2.10)
\begin{equation}
\rho_f(\theta \phi, \vec{P}) = {1 \over{2}} (I^0 (\theta \phi, \vec{P}) \sigma^0 + {\vec{I}} (\theta \phi, \vec{P}) \vec{\sigma})
\end{equation}
\noindent
where the traces $I^j(\theta \phi, \vec{P}) = Tr(\sigma^j \rho_f(\theta \phi, \vec{P})), j=0,1,2,3$ represent measurable intensities of angular distributions as seen from (2.7). Introducing recoil nucleon polarization vector $\vec{Q} (\theta \phi, \vec{P})$ using a relation
\begin{equation}
\vec{Q} (\theta \phi, \vec{P}) I^0(\theta \phi, \vec{P}) \equiv \vec{I} (\theta \phi, \vec{P})
\end{equation}
\noindent
we can write 
\begin{equation}
\rho_f(\theta \phi, \vec{P}) = {1 \over{2}} (1+\vec{Q}(\theta \phi, \vec{P}) \vec{\sigma}) I^0 (\theta \phi, \vec{P})= \rho_n(\vec {Q}) I^0 (\theta \phi, \vec{P}) 
\end{equation}
\noindent
The normalized final state density matrix $\rho^{'}_f(\theta \phi, \vec{P}) =\rho_f(\theta \phi, \vec{P})/{I^0(\theta \phi, \vec{P})}$ is simply the spin density matrix of the recoil nucleon $\rho_n(\vec {Q})$. It will represent a pure final state if and only if the recoil nucleon polarization vector $\vec{Q} = (Q^1,Q^2,Q^3)$ satisfies the condition $|\vec{Q}|^2=1$ for all solid angles $\Omega = (\theta, \phi)$ at any given dipion mass $m$ and momentum transfer $t$~\cite{leader01,nielsen00}.\\

We can see that the vector $\vec Q(\theta \phi, \vec {P})$ defined by (2.12) is recoil nucleon polarization from the definition of polarization vector~\cite{perl74,leader01}. A spin state  of an ensemble of particles with spin posseses a vector polarization when in the rest frame of the particle the spin operator $\vec{s}$ has a non-zero expectation 
value $<\vec{s}> = Tr(\vec{s}\rho)$. In general, a polarization vector $\vec{Q}$ is defined as $\vec{Q} = <\vec{s}>/(s Tr(\rho))=Tr( \vec{s} \rho)/(s Tr( \rho))$. For the final state ensemble of recoil nucleons $\vec{s} = {1 \over{2}} \vec{\sigma}$ and $\rho = \rho_f$, and (2.12) holds.\\

The polarization vector $\vec{Q}$ has transverse components $Q^2$ and $Q^1$ where $Q^2$ is perpendicular to the scattering plane in the direction of the $y$ axis and $Q^1$ is transverse to the neutron direction in the scattering plane. The longitudinal component  $Q^3$ is along the direction of neutron motion in its rest frame. Measurements of $Q^2$ and $Q^1$ can be done by rescattering of the recoil neutron on spin zero carbon target. Measurements of $Q^3$ are more difficult since they require spin rotation by special magnetic fields to convert the longitudinal polarization to transverse polarization.

\subsection{Angular expansion of the final state density matrix.}

Using $\rho_p={1 \over {2}}(1+\vec{P}\vec{\sigma})$ we can write matrix elements of $\rho_f$ in terms of components of target polarization
\begin{equation}
(\rho_f(\theta \phi, \vec{P}))^{{1 \over{2}}{1 \over{2}}}_{\chi \chi'}= (\rho_u(\theta\phi))^{{1 \over{2}}{1 \over{2}}}_{\chi\chi^{'}}+P_x(\rho_x(\theta \phi))^{{1 \over{2}}{1 \over{2}}}_{\chi\chi^{'}}+P_y(\rho_y(\theta \phi))^{{1 \over{2}}{1 \over{2}}}_{\chi \chi^{'}}+P_z(\rho_z(\theta \phi))^{{1 \over{2}}{1 \over{2}}}_{\chi \chi^{'}}
\end{equation}
In (2.14) the subscript $u$ stands for unpolarized target $\vec{P} = 0$. Using the decomposition (2.14) of $\rho_f(\theta \phi, \vec{P})$  we find a decomposition for the intensities
\begin{equation}
I^j(\theta \phi, \vec{P}) = Tr(\sigma^j \rho_f(\theta \phi, \vec{P}))=I^j_u(\theta \phi) + P_x I^j_x(\theta \phi)+ P_y I^j_y(\theta \phi)+ P_z I^j_z(\theta \phi)
\end{equation}
where the components $I^j_k(\theta \phi)$, $j=0,1,2,3$ and $k=u,x,y,z$ of the intensities $I^j(\theta \phi, \vec{P})$ are given by traces
\begin{equation}
I^j_k(\theta \phi) = Tr_{\chi, \chi''} ((\sigma^j)_{\chi \chi''} (\rho_k(\theta \phi))^{{1 \over{2}}{1 \over{2}}}_{\chi'' \chi})
\end{equation}
In general, a plane wave helicity state of two particles with helicities $\mu_1, \mu_2$ can be expanded in terms of angular helicity states ~\cite{perl74,martin70}
\begin{equation}
|p\theta \phi;\mu_1 \mu_2>=\sum \limits_{J,\lambda} \sqrt{{2J+1} \over {4\pi}} D^J_{\lambda,\mu}(\phi,\theta,-\phi)|pJ\lambda;\mu_1 \mu_2>
\end{equation}
where p is the momentum in center-of-mass system and $J$ and $\lambda$ are the two-particle spin and helicity, and $\mu=\mu_1-\mu_2$. For two pions $\mu_1=\mu_2=0$ and $D^J_{\lambda0}(\phi,\theta,-\phi)=\sqrt{4\pi/(2J+1)}Y^{J*}_\lambda (\theta,\phi)$. The final state can be expanded in spherical harmonics
\begin{equation}
|\theta \phi, \chi> = \sum \limits_{J \lambda} Y^{J*}_{\lambda}(\theta, \phi) |J \lambda, \chi>
\end{equation}
\noindent
where $J$ and $\lambda$ are dipion spin and helicity, respectively. Using (2.18), the angular expansion of the transition amplitudes
\begin{equation}
H_{\chi,0\nu}(\theta\phi)=\sqrt{{q(m^2)G(s)}\over {Flux(s)}}T_{\chi,0\nu}(\theta\phi) = \sum \limits_{J \lambda} Y^J_{\lambda}(\theta,\phi) H^J_{\lambda \chi,0\nu}
\end{equation}
\noindent 
defines helicity amplitudes of definite dipion spin $H^J_{\lambda \chi, 0 \nu}(s,t,m)$.
The partial wave helicity amplitudes (2.19) are normalized such that the intensity of $\pi^- \pi^+$ production measured on unpolarized target and integrated over the angular distribution of the produced pions is given by
\begin{equation}
{d^2 \sigma \over{dtdm}} \equiv \int d\Omega I^0_u(\theta\phi)=\int d \Omega Tr_{\chi = \chi^{'}} ((\rho_u(\theta, \phi))^{{1 \over{2}}{1 \over{2}}}_{\chi \chi^{'}})= {1 \over{2}} \sum \limits_{J \lambda} \sum \limits_{\chi, \nu} |H^J_{\lambda \chi,0 \nu}|^2
\end{equation}

The target polarization components of matrix elements (2.14) are given by (2.7)
\begin{equation} 
(\rho_k(\theta\phi))^{{1 \over{2}}{1 \over{2}}}_{\chi\chi^{'}} = {1 \over{2}}\sum \limits_{\nu\nu^{'}}H_{\chi,0\nu}(\theta\phi)(\sigma_k)_{\nu\nu^{'}}H^*_{\chi^{'},0\nu^{'}}
(\theta\phi)
\end{equation}
\noindent
where $k=u,x,y,z$ and $(\sigma_u)_{\nu\nu^{'}}=\delta_{\nu\nu^{'}}$ is a unit matrix. Using (2.19) their angular expansion reads
\begin{equation}
(\rho_k(\theta \phi))^{{1 \over{2}}{1 \over{2}}}_{\chi \chi^{'}}= \sum \limits_{J \lambda} \sum  \limits_{J^{'} \lambda^{'}} (R_k)^{J {1 \over{2}}, J^{'} {1 \over{2}}}_{\lambda \chi, \lambda^{'} \chi^{'}} Y^{J}_{\lambda}(\theta,\phi) Y^{J^{'*}}_{\lambda^{'}}(\theta, \phi)
\end{equation}
\noindent
where $(R_k)^{J {1 \over{2}}, J^{'} {1 \over{2}}}_{\lambda \chi, \lambda^{'} \chi^{'}}$ are the angular spin density matrix elements of the final density matrix. In terms of partial wave helicity amplitudes $H^J_{\lambda\chi,0\nu}$ these angular matrix elements read
\begin{equation}
(R_k(s,t,m))^{J{1 \over{2}}, J^{'} {1 \over{2}}}_{\lambda\chi, \lambda^{'} \chi^{'}}=
{1 \over{2}} \sum \limits_{\nu \nu^{'}} H^J_{\lambda \chi ,0 \nu} (\sigma_k)_{\nu \nu^{'}} H^{J^{'}*}_{\lambda^{'} \chi^{'},0 \nu^{'}}\equiv (\rho_k(s,t,m))^{J{1 \over{2}}, J^{'} {1 \over{2}}}_{\lambda\chi, \lambda^{'} \chi^{'}}{{d^2 \sigma} \over{dtdm}}
\end{equation}
\noindent
where we defined normalized angular spin density matrix elements $(\rho_k(s,t,m))^{J{1 \over{2}}, J^{'} {1 \over{2}}}_{\lambda\chi, \lambda^{'} \chi^{'}}$. The component intensities $I^j_k(\theta \phi)$, $k=u,x,y,z$, $j=0,1,2,3$ have angular expansions arising from the traces (2.16)
\begin{equation}
I^j_k(\theta \phi) \equiv (\rho^j_k(\theta\phi)){{d^2 \sigma} \over{dtdm}} =Tr_{\chi,\chi^{''}}((\sigma^j)_{\chi\chi^{''}}(\rho_k(\theta \phi))^{{1 \over{2}}{1 \over{2}}}_{\chi'' \chi})=\sum \limits_{J \lambda} \sum \limits_{J' \lambda'} (R^j_k)^{JJ'}_{\lambda \lambda'} Y^J_{\lambda}(\theta \phi) Y^{J'*}_{\lambda'}(\theta \phi)
\end{equation}
\noindent
where the unnormalized angular density matrix elements $R^j_k$ are the traces over recoil nucleon helicities
\begin{equation}
(R^j_k(s,t,m))^{JJ'}_{\lambda \lambda'}= Tr_{\chi, \chi''}((\sigma^j)_{\chi\chi''}(R_k)^{J{1 \over{2}}, J^{'} {1 \over{2}
}}_{\lambda\chi'', \lambda^{'} \chi})
\end{equation}
\noindent 
Expressed in terms of partial wave helicity amplitudes they read
\begin{equation}
(R^j_k)^{JJ'}_{\lambda \lambda'} \equiv (\rho^j_k)^{JJ'}_{\lambda \lambda'} {{d^2 \sigma} \over{dtdm}}= {1 \over{2}} \sum \limits_{\chi,\chi''} \sum \limits_{\nu \nu'} (\sigma^j)_{\chi\chi''}H^J_{\lambda \chi'', 0 \nu}(\sigma_k)_{\nu \nu'} H^{J'*}_{\lambda' \chi, 0 \nu'}
\end{equation}
where we defined normalized angular density matrix elements 
$(\rho^j_k)^{JJ'}_{\lambda \lambda'}$. The normalization conditions
\begin{equation}
Tr_{J=J^{'},\lambda=\lambda^{'}}(Tr_{\chi=\chi^{'}} (\rho_u(s,t,m))^{J{1 \over{2}}, J^{'} {1 \over{2}}}_{\lambda\chi, \lambda^{'}\chi^{'}})
=Tr_{J=J^{'},\lambda=\lambda^{'}}((\rho^0_u(s,t,m))^{J,J{'}}_{\lambda, \lambda^{'}})=1
\end{equation}
\noindent
are equivalent to (2.20) for normalized helicity amplitudes.\\ 

Combining in the sum (2.25) the terms with inverted $J \lambda$ and $J' \lambda'$ we can write (2.25) for each $k=u,y,x,y$ and $j=0,1,2,3$ in the form
\begin{equation}
{1 \over{4}} \sum \limits_{J \lambda} \sum \limits_{J' \lambda'} 
[R^{JJ'}_{\lambda \lambda'} Y^J_{\lambda} Y^{J'*}_{\lambda'} + 
R^{JJ'}_{-\lambda -\lambda'} Y^J_{-\lambda} Y^{J'*}_{-\lambda'} + 
R^{J'J}_{\lambda' \lambda} Y^{J'}_{\lambda'} Y^{J*}_{\lambda} + 
R^{J'J}_{-\lambda' -\lambda} Y^{J'}_{-\lambda'} Y^{J*}_{-\lambda}]
\end{equation}
\noindent
Using hermiticity of the density matrix
\begin{equation}
(R^j_k)^{J'J}_{\lambda' \lambda}=(R^j_k)^{JJ'*}_{\lambda \lambda'}
\end{equation}
\noindent
and a relation for spherical harmonics $Y^L_{-M} (\theta, \phi) = (-1)^M (Y^L_{M} (\theta, \phi))^{*}$ the sum of terms in (2.29) takes the form
\begin{equation} 
[+2Re(R^{JJ'}_{\lambda \lambda'} + (-1)^{\lambda+ \lambda'} R^{J J' *}_{- \lambda -\lambda'})
Re(Y^J_\lambda Y^{J'*}_{\lambda'})
\end{equation}
\[
 -2Im(R^{JJ'}_{\lambda \lambda'} + (-1)^{\lambda+ \lambda'} R^{J J' *}_{- \lambda -\lambda'})
Im(Y^J_\lambda Y^{J'*}_{\lambda'})]
\]
The amplitudes $H^J_{\lambda \chi, 0 \nu}$ describe a two-body process $\pi^- + p \to "J(m^2)" + n$ where $J$ and $m^2$ are the spin and mass of the dipion particle $"J(m^2)"$. The initial and final states in this process are separable. For such states the parity conservation in strong interactions requires that~\cite{martin70,perl74,leader01}
\begin{equation}
H^J_{-\lambda - \chi,0 - \nu} = (-1)^{\lambda+ \chi + \nu} H^J_{\lambda \chi, 0 \nu}
\end{equation}
Parity conservation relations (2.32) imply symmetry relations for spin density matrix elements
\begin{equation}
(R^j_k)^{JJ'}_{\lambda \lambda'} = + (-1)^{\lambda + \lambda'} (R^j_k)^{JJ'}_{-\lambda -\lambda'} 
\end{equation}
\noindent
for $(k,j)=(u,0),(y,0),(u,2),(y,2),(x,1),(z,1),(x,3),(z,3)$ and
\begin{equation}
(R^j_k)^{JJ'}_{\lambda \lambda'} = - (-1)^{\lambda + \lambda'} (R^j_k)^{JJ'}_{-\lambda -\lambda'} 
\end{equation}
\noindent
for $(x,0),(z,0),(x,2),(z,2),(u,1),(y,1),(u,3),(y,3)$. Using these symmetry relations the components $I^j_k(\theta \phi)$ of the dipion angular distribution $I^j(\theta \phi, \vec{P})$ measured on polarized target take the form
\begin{equation}
I^j_k(\theta \phi)=\sum \limits_{J \lambda} \sum \limits_{J' \lambda'} (Re R^j_k)^{JJ'}_{\lambda \lambda'} Re(Y^J_\lambda(\theta \phi)Y^{J'*}_{\lambda'}(\theta \phi)) = {{d^2 \sigma} \over{dtdm}} \sum \limits_{J \lambda} \sum \limits_{J' \lambda'} (Re \rho^j_k)^{JJ'}_{\lambda \lambda'} Re(Y^J_\lambda(\theta \phi)Y^{J'*}_{\lambda'}(\theta \phi))
\end{equation}
\noindent
for $(k,j)=(u,0),(y,0),(u,2),(y,2),(x,1),(z,1),(x,3),(z,3)$ and
\begin{equation}
I^j_k(\theta \phi)=\sum \limits_{ J \lambda} \sum \limits_{J' \lambda'} (Im R^j_k)^{JJ'}_{\lambda \lambda'} Im(Y^J_\lambda(\theta \phi)Y^{J'*}_{\lambda'}(\theta \phi)) = {{d^2 \sigma} \over{dtdm}} \sum \limits_{ J \lambda} \sum \limits_{J' \lambda'} (Im \rho^j_k)^{JJ'}_{\lambda \lambda'} Im(Y^J_\lambda(\theta \phi)Y^{J'*}_{\lambda'}(\theta \phi))
\end{equation}
\noindent
for $(x,0),(z,0),(x,2),(z,2),(u,1),(y,1),(u,3),(y,3)$. The elements 
$(Im R^j_k)^{JJ'}_{\lambda \lambda'}$ in the group (2.34) and $(Re R^j_k)^{JJ'}_{\lambda \lambda'}$ in the group (3.35) are not observable as the result of parity conservation in strong interactions. However, as we shall see in Section ?? in the case of $S$- and $P$-waves, the measured elements on polarized targets supplemented by certain phase relations between amplitudes enable to calculate the unobservable elements. 

\subsection{Experimental form of angular distributions.}

The expressions (2.34) and (2.35) for angular intensities $I^j_k(\theta \phi)$ in terms of angular matrix elements $(R^j_k)^{JJ'}_{\lambda \lambda'}$ can be simplified. Using the hermiticity of the spin density matrix elements $(\rho^j_k)^{J'J}_{\lambda' \lambda}=(\rho^j_k)^{JJ'*}_{\lambda \lambda'}$ we can write (2.34) and (2.35) in the form
\begin{equation}
I^j_k(\theta \phi) = {{d^2 \sigma} \over{dtdm}}
[\sum \limits_{J} \sum \limits _{\lambda \lambda'} (Re \rho^j_k)^{JJ}_{\lambda \lambda'} Re(Y^J_\lambda(\theta \phi)Y^{J*}_{\lambda'}(\theta \phi))
+\sum \limits_{J < J'} \sum \limits_{\lambda \lambda'} 2(Re \rho^j_k)^{JJ'}_{\lambda \lambda'} Re(Y^J_\lambda(\theta \phi)Y^{J'*}_{\lambda'}(\theta \phi))]
\end{equation}
\noindent
\begin{equation}
I^j_k(\theta \phi) = {{d^2 \sigma} \over{dtdm}} \sum \limits_{ J < J'} \sum \limits_{\lambda \lambda'} 2(Im \rho^j_k)^{JJ'}_{\lambda \lambda'} Im(Y^J_\lambda(\theta \phi)
Y^{J'*}_{\lambda'}(\theta \phi))
\end{equation}
\noindent
The number of spin density matrix elements is further reduced by making use of their parity symmetry relations (2.32) and (2.33) and relations for spherical harmonics
\begin{equation}
Re(Y^J_{-\lambda}Y^{J'*}_{-\lambda'})=(-1)^{\lambda+\lambda'}Re(Y^J_{\lambda}Y^{J'*}_{\lambda'})
\end{equation}
\[
Im(Y^J_{-\lambda}Y^{J'*}_{-\lambda'})=-(-1)^{\lambda+\lambda'}Im(Y^J_{\lambda}Y^{J'*}_{\lambda'})
\]
We can then simplify the sums
\begin{equation}
\sum \limits_{\lambda, \lambda'} =
\sum \limits_{\lambda \geq 0} \sum_{\lambda'}+ \sum \limits_{\lambda < 0} \sum_{\lambda'}=
\sum \limits_{\lambda \geq 0} \sum_{\lambda'} \xi_\lambda (...)
\end{equation}
where $\xi_0=1$ and $\xi_\lambda=2$ for $\lambda >0$. The two sums in (2.36) can be combined by introducing a factor $\xi_{JJ'}=1$ for $J=J'$ and $\xi_{JJ'}=2$ for $J<J'$. The final expressions for (2.34) and (2.35) with independent angular density matrix elements then read
\begin{equation}
I^j_k(\theta \phi) = {{d^2 \sigma} \over{dtdm}}
\sum \limits_{J \leq J'} \sum \limits _{\lambda \geq 0} \sum \limits_{\lambda'} 
\xi_{JJ'} \xi_\lambda (Re \rho^j_k)^{JJ}_{\lambda \lambda'} Re(Y^J_\lambda(\theta\phi)Y^{J*}_{\lambda'}(\theta \phi))
\end{equation}
\noindent
\begin{equation}
I^j_k(\theta \phi) = {{d^2 \sigma} \over{dtdm}} 
\sum \limits_{ J \leq J'} \sum \limits_{\lambda \geq 0} \sum \limits_{\lambda'} 
\xi_{JJ'} \xi_\lambda (Im \rho^j_k)^{JJ'}_{\lambda \lambda'} 
Im(Y^J_\lambda(\theta \phi)Y^{J'*}_{\lambda'}(\theta \phi))
\end{equation}
Note that in (2.41) $Im(\rho^j_k)^{JJ}_{\lambda \lambda}=0$ and $Im(Y^J_0(\theta \phi)Y^{J'*}_0(\theta \phi))=0$. Because of the angular properties of $Y^1_\lambda(\theta \phi)$, the elements $(\rho^j_k)^{00}_{00} \equiv (\rho^j_k)^{00}_{ss}$, $(\rho^j_k)^{11}_{00}$ and $(\rho^j_k)^{11}_{11}$ are not independent but appear in two independent combinations 
\begin{equation}
(\rho^j_k)_{SP} \equiv (\rho^j_k)^{00}_{ss}+(\rho^j_k)^{11}_{00}+2(\rho^j_k)^{11}_{11}, 
\qquad
(\rho^j_k)_{PP} \equiv (\rho^j_k)^{11}_{00}-(\rho^j_k)^{11}_{11}
\end{equation}

The most feasible experiments are measurements on unpolarized or polarized targets which measure the two-pion angular distribution $I^0(\theta \phi, \vec{P})$ leaving the recoil nucleon polarization vector $\vec{Q}$ not observed. Such measurements provide information on the reduced final state density matrix given by $I^0(\theta \phi, \vec{P})$ 
\begin{equation}
I^0(\theta \phi,\vec{P}) = Tr_{\chi = \chi'} ((\rho_f(\theta \phi,\vec{P})^{{1 \over{2}} 
{1 \over{2}}}_{\chi \chi'})=I^0_u(\theta \phi)+P_x I^0_x(\theta \phi) + P_y I^0_y(\theta \phi) + P_z I^0_z(\theta \phi)
\end{equation}
It is the intensity $I^0(\theta \phi, \vec{P})$ which has been measured in CERN measurements of pion creation processes on transversely polarized targets and on which this study is based.\\

Experimentally, in a given mass region only helicity amplitudes with $J \leq J_{max}$ contribute and all sums in (2.40) and (2.41) are finite. The spin observables $(\rho^j_k)^{JJ'}_{\lambda \lambda'}$ for $J \leq J_{max}$ are determined in small $(m,t)$ bins from the measured angular distribution $I^j(\theta \phi, \vec{P})$ at any given target polarization $\vec{P}$ using the statistical methods of maximum likelihood function~\cite{eadie71,grayer74,becker79a,lesquen85}. This process is referred to as quantum state tomography in Quantum Information Theory~\cite{nielsen00}.

\subsection{Nucleon helicity and transversity amplitudes with definite $t$-channel naturality.}

The helicity amplitudes $H^J_{\lambda \chi,0 \nu}$ are combinations of helicity amplitudes with definite $t$-channel naturality $\eta = \mathcal{P} \mathcal{S}$
where $\mathcal{P}$ and $\mathcal{S}$ are the parity and the signature of Reggeons exchanged in $\pi^- + p \to "J(m^2)" + n$~\cite{martin70}. The natural and unnatural amplitudes $N^J_{\lambda +,0 \pm}$ and $U^J_{\lambda +,0 \pm}$  correspond to naturality $\eta = +1$ and $\eta = -1$, respectively. They are given for $\lambda \neq 0$ by relations~\cite{lutz78,bourrely80,leader01}
\begin{equation}
U^J_{\lambda +,0 \pm} = {1 \over{\sqrt{2}}} (H^J_{\lambda +,0 \pm} + 
(-1)^\lambda H^J_{- \lambda +,0 \pm}) 
\end{equation}
\[
N^J_{\lambda +,0 \pm} = {1 \over{\sqrt{2}}} (H^J_{\lambda +,0 \pm} - 
(-1)^\lambda H^J_{- \lambda +,0 \pm}) 
\]
For $\lambda = 0$ they are 
\begin{equation}
U^J_{0+,0 \pm} = H^J_{0+,0+} \text{  and } N^J_{0+,0 \pm} = 0
\end{equation}  
In (2.44) and (2.45) + and - correspond to $+{1 \over {2}}$ and $-{1 \over {2}}$ values of nucleon helicities. The unnatural exchange amplitudes $U^J_{\lambda +,0-}$ and $U^J_{\lambda +,0+}$ exchange $\pi$ and $a_1$ quantum numbers in the $t$-channel, respectively, while the natural exchange amplitudes $N^J_{\lambda +,0-}$ and $N^J_{\lambda +,0+}$ both exchange $a_2$ quantum numbers.\\

Amplitude analyses of measurements on polarized targets are best performed in terms of transversity amplitudes with definite $t$-channel naturality~\cite{lutz78,bourrely80,leader01}. In such measurements the spin states of the target nucleon are described by transversity $\tau$ with $\tau = +{1 \over{2}} \equiv u$ and $\tau = -{1 \over{2}} \equiv d$ corresponding to "up" and "down" orientations of the target spin relative to the scattering plane~\cite{bourrely80,leader01}. Following Lutz and Rybicki~\cite{lutz78}, we define mixed helicity-transversity amplitudes with nucleon helicity replaced by nucleon transversity
\begin{equation}
G^J_{\lambda \tau_n,0 \tau_p} = \sum\limits_{\lambda_p,\lambda_n} \xi D^{{1\over{2}}*}_{\tau_n \lambda_n }({\pi \over{2}}, {\pi \over{2}},-{\pi \over{2}})H^J_{\lambda \lambda_n,0 \lambda_p}D^{1\over{2}}_{\lambda_p \tau_p}({\pi \over{2}}, {\pi \over{2}},-{\pi \over{2}})
\end{equation}
\[
\text{  where  } D^{1\over{2}}_{\lambda \tau}({\pi \over{2}}, {\pi \over{2}}, -{\pi \over{2}})=e^{i {\pi \over{2}} (\lambda - \tau)} d^{1\over{2}}_{\lambda \tau}({\pi \over{2}})
\]
The factor $\xi=1$ or $\xi=e^{i \pi (\lambda_n - \lambda_p)}$ for the $y$ axis in the rest frames of target and recoil nucleon in the direction or in opposite direction to the normal to the scattering plane, respectively~\cite{lesquen89,bourrely80,leader01}. In our case $\xi=1$. Then for $\lambda=0$ and $\lambda \neq 0$ we obtain
\begin{equation}
G^J_{0d,0u} = -i(H^J_{0 +,0+} + iH^J_{0 +,0-}), \qquad
G^J_{0u,0d} = +i(H^J_{0 +,0+} - iH^J_{0 +,0-})
\end{equation}
\[
G^J_{\lambda u,0u} = H^J_{\lambda +,0+} + iH^J_{\lambda +,0-}, \qquad
G^J_{\lambda d,0d} = H^J_{\lambda +,0+} - iH^J_{\lambda +,0-}
\]
Parity conservation (2.31) requires that
\begin{equation}
G^J_{0u,0u}=G^J_{0d,0d}=G^J_{\lambda d,0u}=G^J_{\lambda u,0d}=0
\end{equation}
Omitting the inessential factors $\pm i$ in front of parenthesis in (2.47), the unnatural exchange transversity amplitudes are given by
\begin{equation}
U^J_\lambda \equiv U^J_{\lambda,u} = U^J_{\lambda \tau_n,0u} = 
{1 \over{\sqrt{2}}} (G^J_{\lambda \tau_n,0 u}+(-1)^\lambda G^J_{- \lambda \tau_n,0 u}) =
{1 \over{\sqrt{2}}} (U^J_{\lambda +,0+} + i U^J_{\lambda +,0-})
\end{equation}
\[
\overline{U}^J_\lambda \equiv U^J_{\lambda,d} = U^J_{\lambda \tau_n,0d} = 
{1 \over{\sqrt{2}}} (G^J_{\lambda \tau_n,0 d}+(-1)^\lambda G^J_{- \lambda \tau_n,0 d}) =
{1 \over{\sqrt{2}}} (U^J_{\lambda +,0+} - i U^J_{\lambda +,0-})
\]
where $\tau_n=-\tau_p$ and $\tau_n=+\tau_p$ for $\lambda=0$ and $\lambda \neq 0$, respectively. The natural exchange transversity amplitudes are given by
\begin{equation}
N^J_\lambda \equiv N^J_{\lambda,u} = N^J_{\lambda u,0u} = 
{1 \over{\sqrt{2}}} (G^J_{\lambda u,0 u}-(-1)^\lambda G^J_{- \lambda u,0 u}) = 
{1 \over{\sqrt{2}}} (N^J_{\lambda +,0+} + i N^J_{\lambda +,0-})
\end{equation}
\[
\overline{N}^J_\lambda \equiv N^J_{\lambda,d} = N^J_{\lambda d,0d} = 
{1 \over{\sqrt{2}}} (G^J_{\lambda d,0 d}-(-1)^\lambda G^J_{- \lambda d,0 d}) = 
{1 \over{\sqrt{2}}} (N^J_{\lambda +,0+} - i N^J_{\lambda +,0-})
\]
\[
N^J_0 = \overline{N}^J_0 = 0 \text{  or  } N^J_{0,u} = N^J_{0,d} = 0 
\]
The factor $1/\sqrt{2}$ results in the normalization 
\begin{equation} 
{d^2 \sigma \over{dtdm}} \equiv \int d\Omega I^0_u(\theta\phi)=\sum \limits_{J \lambda} \sum \limits_{\tau_p} |U^J_{\lambda,\tau_p}|^2 + |N^J_{\lambda, \tau_p}|^2
\end{equation}
Parity conservation requires 
\begin{equation}
U^J_{\lambda}=+(-1)^\lambda U^J_\lambda, \qquad 
\overline{U}^J_{\lambda}=+(-1)^\lambda \overline{U}^J_\lambda
\end{equation}
\[
N^J_{\lambda}=-(-1)^\lambda N^J_\lambda, \qquad 
\overline{N}^J_{\lambda}=-(-1)^\lambda \overline{N}^J_\lambda
\]
Inverted relations for helicity amplitudes in terms of transversity amplitudes with definite $t$-channel naturality for $\lambda \geq 0$ read
\begin{equation}
H^J_{\lambda+,0+}={1 \over{2}} 
[(U^J_\lambda + \overline{U}^J_\lambda) + (N^J_\lambda + \overline{N}^J_\lambda)]
\end{equation}
\[
H^J_{- \lambda+,0+}={{(-1)^\lambda} \over{2}} 
[(U^J_\lambda + \overline{U}^J_\lambda) - (N^J_\lambda + \overline{N}^J_\lambda)]
\]
\[
H^J_{\lambda+,0-}={{-i} \over{2}} 
[(U^J_\lambda - \overline{U}^J_\lambda) + (N^J_\lambda - \overline{N}^J_\lambda)]
\]
\[
H^J_{- \lambda+,0-}={{-i(-1)^\lambda} \over{2}} 
[(U^J_\lambda - \overline{U}^J_\lambda) - (N^J_\lambda - \overline{N}^J_\lambda)]
\]

General expressions for all spin observables $(R^j_k)^{JJ'}_{\lambda \lambda'}$ for several types of helicity and transversity amplitudes were calculated by Lutz and Rybicki and are given in their unpublished paper Ref.~\cite{lutz78}. Their original Tables 1a, 1d and 1b for helicity amplitudes, transversity amplitudes with definite $t$-channel naturality and helicity amplitudes with definite $t$-channel naturality, respectively, are reproduced in the Appendix A.

\subsection{Trace measurements.}

The traces $Tr(Re \rho^j_k)$ can be measured in experiments on unpolarized or polarized targets with or without measurements of recoil nucleon polarization in which the angular distribution of the produced pions is not observed. The traces are given by the integrated angular distributions
\begin{equation}
{\overline{I}}^j_k = \int I^j_k(\Omega) d \Omega = {{d^2 \sigma} \over{dtdm}}
\sum \limits_{J \lambda} (Re \rho^j_k)^{JJ}_{\lambda \lambda} = {{d^2 \sigma} \over{dtdm}} Tr(Re \rho^j_k)
\end{equation}
for $(k,j)=(u,0),(y,0),(u,2),(y,2),(x,1),(z,1),(x,3),(z,3)$. The integrated angular distributions 
\begin{equation}
{\overline{I}}^j_k= \int I^j_k (\theta \phi) d \Omega = 0
\end{equation}
vanish for $(k,j)=(x,0),(z,0),(x,2),(z,2),(u,1),(y,1),(u,3),(y,3)$ since the integral
\begin{equation}
\int Y^J_\lambda (\theta \phi) Y^{J'*}_{\lambda'} (\theta \phi) d \Omega = {\delta}_{JJ'} {\delta}_{\lambda \lambda'}
\end{equation}
is real~\cite{edmonds57}. The intensities $I^0_k(\theta \phi)$ provide information about the polarized target asymetry vector $\vec{T} = (T_x, T_y, T_z)$ defined by
\begin{equation}
T_k(\theta \phi)I^0_u(\theta \phi) \equiv I^0_k(\theta \phi)
\end{equation}
The averaged values of its components
\begin{equation}
{\overline{T}}_k ={{\int T_k(\theta \phi) I^0_u(\theta \phi) d \Omega} \over{ \int I^0_u (\theta \phi) d \Omega}} = {{\int I_k(\theta \phi) d \Omega} \over{\int I^0_u (\theta \phi) d \Omega}}
\end{equation}
are $\overline{T}_x = \overline{T}_z =0$ and $\overline{T}_y \equiv T$ where $T$ is polarized target asymmetry. With
\begin{equation}
{{d^2 \sigma} \over{dtdm}} = \int I^0_u(\theta \phi) d \Omega \text{  and  } T{{d^2 \sigma} \over{dtdm}} = \int I^0_y (\theta \phi) d \Omega
\end{equation}
we get from (2.54) trace conditions
\begin{equation}
\sigma \equiv \sum \limits_{J \lambda} (Re \rho^0_u)^{JJ}_{\lambda \lambda} =1 \text{  and  } T = \sum \limits_{J \lambda} (Re \rho^0_y)^{JJ}_{\lambda \lambda} 
\end{equation}
When the recoil nucleon polarization is not measured, the integrated intensity measured on polarized target 
\begin{equation}
{\overline{I}}^0(\vec{P})={\overline{I}}^0_u+P_y {\overline{I}}^0_y=(1 + P_y T){{d^2 \sigma} \over{dtdm}}
\end{equation}
does not depend on components $P_x$ and $P_z$ of the target polarization.\\

When the $\pi^- \pi^+$ or $\pi^0 \pi^0$ angular distribution is not observed, we measure the average values $\overline{Q}^j(\vec{P})$ of the components of the recoil nucleon polarization vector
\begin{equation}
\overline{Q}^j(\vec{P}) = {{\int Q^j(\theta \phi) I^0(\theta \phi) d \Omega} \over{\int I^0(\theta \phi) d \Omega}} = {{\int I^j(\theta \phi) d \Omega} \over{\int I^0(\theta \phi) d \Omega}}
\end{equation}
Using (2.15) we can write
\begin{equation}
\overline{Q}^j(\vec{P}) \overline{I}^0 = \overline{I}^j = \overline{I}^j_u + 
P_x \overline{I}^j_x + P_y \overline{I}^j_y + P_z \overline{I}^j_z
\end{equation}
With averaged intensities (2.55) vanishing due to parity conservation, we have the following result for the recoil nucleon polarization four-vector $\overline{Q}^j, j=0,1,2,3$
\begin{equation}
\overline{Q}^0(\vec{P}) \overline{I}^0(\vec{P}) = \overline{I}^0_u + P_y \overline{I}^0_y
=(\overline{Q}^0_u + P_y \overline{Q}^0_y){{d^2 \sigma} \over{dtdm}}
\end{equation}
\[
\overline{Q}^2(\vec{P}) \overline{I}^0(\vec{P}) = \overline{I}^2_u + P_y \overline{I}^2_y
=(\overline{Q}^2_u + P_y \overline{Q}^2_y){{d^2 \sigma} \over{dtdm}}
\]
\[
\overline{Q}^1(\vec{P}) \overline{I}^0(\vec{P}) = P_x \overline{I}^1_x + P_z \overline{I}^1_z
=(P_x \overline{Q}^1_x + P_z \overline{Q}^1_z){{d^2 \sigma} \over{dtdm}}
\]
\[
\overline{Q}^3(\vec{P}) \overline{I}^0(\vec{P}) = P_x \overline{I}^3_x + P_z \overline{I}^3_z
=(P_x \overline{Q}^3_x + P_z \overline{Q}^3_z){{d^2 \sigma} \over{dtdm}}
\]
For transversely polarized target with $P_x=P_z=0$ only the transverse component 
$\overline {Q}^2$ of the averaged recoil nucleon polarization is non-zero. Note that $\overline{Q}^0(\vec{P})=1$.

\subsection{The structure of the averaged recoil nucleon polarization vector.}

Using the expressions (2.26) and the parity relations (2.31) we can express the traces $\overline{I}^j_k$ in terms of helicity amplitudes. We find
\begin{equation}
\overline{I}^0_u = \sigma {{d^2 \sigma} \over{dtdm}} = \sum \limits_{J \lambda} 
(|H^J_{\lambda +,0+}|^2 + (|H^J_{\lambda +,0-}|^2 )
\end{equation}
\[
\overline{I}^0_y = T {{d^2 \sigma} \over{dtdm}} = \sum \limits_{J \lambda}
2Im(H^J_{\lambda +,0+} H^{J*}_{\lambda +,0-} )
\]
\[
\overline{I}^3_x = A {{d^2 \sigma} \over{dtdm}} = \sum \limits_{J \lambda} 
2Re(H^J_{\lambda +,0+} H^{J*}_{\lambda +,0-} )
\]
\[
\overline{I}^3_z = R {{d^2 \sigma} \over{dtdm}} = \sum \limits_{J \lambda} 
(|H^J_{\lambda +,0+}|^2 - (|H^J_{\lambda +,0-}|^2 )
\]
where $A$ and $R$ are spin rotation parameters defined in analogy with $\pi N \to \pi N$ scattering~\cite{perl74,bourrely80,leader01}. For the remaining traces we get
\begin{equation}
\overline{I}^2_u =\sum \limits_{J \lambda}+(-1)^\lambda Im(H^J_{- \lambda+,0-} H^{J*}_{\lambda+,0+} - H^J_{- \lambda+,0+} H^{J*}_{\lambda+,0-})
\end{equation}
\[
\overline{I}^2_y =\sum \limits_{J \lambda}-(-1)^\lambda Re(H^J_{- \lambda+,0+} H^{J*}_{\lambda+,0+} + H^J_{- \lambda+,0-} H^{J*}_{\lambda+,0-})
\]
\[
\overline{I}^1_x =\sum \limits_{J \lambda}-(-1)^\lambda Re(H^J_{- \lambda+,0+} H^{J*}_{\lambda+,0+} - H^J_{- \lambda+,0-} H^{J*}_{\lambda+,0-})
\]
\[
\overline{I}^1_z =\sum \limits_{J \lambda}+(-1)^\lambda Re(H^J_{- \lambda+,0+} H^{J*}_{\lambda+,0+} + H^J_{- \lambda+,0-} H^{J*}_{\lambda+,0-})
\]
In our next step we shall relate the four traces in (2.66) to the four traces in (2.65). After some algebraic rearrangements of the terms and sums in (2.66) and using parity relations for the helicity amplitudes we find
\begin{equation}
\overline{I}^2_u = (-T + 2 \tau_N){{d^2 \sigma} \over{dtdm}}, \qquad
\overline{I}^2_y = (-\sigma + 2 \sigma_N){{d^2 \sigma} \over{dtdm}}
\end{equation}
\[
\overline{I}^1_x = (- R + 2 R_N){{d^2 \sigma} \over{dtdm}}, \qquad
\overline{I}^1_z = (+ A - 2 A_N){{d^2 \sigma} \over{dtdm}}
\]
For now, the traces $\sigma, T, R, A$ in (2.67) are expressed in terms of helicity amplitudes as shown in (2.65). The new terms $\sigma_N, \tau_N, R_N, A_N$ are expressed in terms of natural exchange helicity amplidudes and are defined next. With $A^J_\lambda=U^J_\lambda,N^J_\lambda$ for unnatural and natural exchange amplitudes, we define normalized partial wave intensities $\sigma_A(J \lambda)$ and normalized partial wave polarizations $\tau_A(J \lambda)$
\begin{equation}
\sigma_A(J \lambda){{d^2 \sigma} \over{dtdm}} \equiv 
|A^J_\lambda|^2 + |\overline{A}^J_\lambda|^2, \qquad
\tau_A(J \lambda){{d^2 \sigma} \over{dtdm}}   \equiv 
|A^J_\lambda|^2 - |\overline{A}^J_\lambda|^2
\end{equation}
Similarly we define partial wave spin rotation parameters
\begin{equation}
A_A(J \lambda){{d^2 \sigma} \over{dtdm}}    \equiv
2Im(A^J_\lambda \overline{A}^{J*}_\lambda), \qquad 
R_A(J \lambda){{d^2 \sigma} \over{dtdm}}    \equiv
2Re(A^J_\lambda \overline{A}^{J*}_\lambda)
\end{equation}
Next we need to define their sums for $\lambda \geq 0$
\begin{equation}
\sigma_A = \sum \limits_{J \lambda \geq 0} \sigma_A(J \lambda), \quad 
\tau_A = \sum \limits_{J \lambda \geq 0} \tau_A(J \lambda),
\end{equation}
\[
A_A = \sum \limits_{J \lambda \geq 0} A_A(J \lambda), \quad 
R_A = \sum \limits_{J \lambda \geq 0} R_A(J \lambda)
\]
When the traces (2.65) and (2.66) are expressed in terms of transversity amplitudes with definite naturality using the relations (2.53), we find 
\begin{equation}
\overline{Q}^0_u=\sigma_U + \sigma_N =\sigma , \qquad \overline{Q}^0_y=\tau_U + \tau_N=T
\end{equation}
\[
\overline{Q}^2_u= -T + 2 \tau_N, \qquad \overline{Q}^2_y = -\sigma + 2 \sigma_N
\]
and
\begin{equation}
\overline{Q}^3_x = A_U +A_N=A, \qquad \overline{Q}^3_z = R_U + R_N=R
\end{equation}
\[
\overline{Q}^1_x= -R + 2R_N, \qquad \overline{Q}^1_z = +A-2A_N
\]
where $\sigma=1$. The equations (2.72) and (2.73) reveal simple structure of the averaged recoil nucleon polarization vector. We make an important observation that the components $\overline{Q}^2_u$ and $\overline{Q}^2_y$ are fully determined by measurements on polarized target without any measurement of recoil nucleon polarization and that these components differ from the traces $\sigma$ and $T$ by natural exchange terms $\sigma_N$ and $\tau_N$. It is this difference that turned out to be crucial in the initial development of experimental tests of unitarity of the $S$-matrix.

\section{Test of unitarity in $\pi N \to \pi N$ and similar two-body processes.}

The final state density matrix in reaction $\pi N \to \pi N$ measured on polarized target with polarization $\vec {P}=(P_x,P_y,P_z)$ has the form
\begin{equation}
\rho_f(\vec {P}) = {1 \over{2}}\bigl (I^0(\vec {P})+\vec {I}(\vec {P})\vec {\sigma}\bigr )=
{1 \over{2}}\bigl (1+\vec {Q}(\vec {P})\vec {\sigma}\bigr )I^0(\vec {P})
\end{equation}
where $\vec {Q}(\vec {P})$ is the recoil nucleon polarization. Assuming parity conservation
for helicity amplitudes $H_{0-\lambda',0-\lambda}=-(-1)^{\lambda+\lambda'}H_{0\lambda',0\lambda}$ and using the definitions  (2.26), we obtain the well known expressions for components $I^j(\vec {P})=Q^j(\vec {P})I^0(\vec {P})$~\cite{delesquen72,perl74}
\begin{equation}
I^0(\vec {P})= Q^0(\vec {P})I^0(\vec {P})= (1+TP_y){d\sigma \over{dt}}
\end{equation}
\[
I^1(\vec {P})= Q^1(\vec {P})I^0(\vec {P})= (RP_x-AP_z){d\sigma \over{dt}}
\]
\[
I^2(\vec {P})= Q^2(\vec {P})I^0(\vec {P})= (T+P_y){d\sigma \over{dt}}
\]
\[
I^3(\vec {P})= Q^3(\vec {P})I^0(\vec {P})= (AP_x+RP_z){d\sigma \over{dt}}
\]
where $Q^0(\vec {P})=1$. The differential cross section ${d\sigma/{dt}}$, polarized target asymmetry $T$ (often called polarization $P$) and spin rotation parameters $A$ and $R$ are defined in terms of helicity or transversity amplitudes as follows
\begin{equation}
{d\sigma \over {dt}}= |H_{0+,0+}|+|H_{0+,0-}|^2=|H_u|^2+|H_d|^2
\end{equation}
\[
T{d\sigma \over {dt}}= 2Im(H_{0+,0+}H_{0+,0-}^*) = |H_u|^2-|H_d|^2
\]
\[
A{d\sigma \over {dt}}= 2Re(H_{0+,0+}H_{0+,0-}^*) = 2Im(H_dH_u^*)
\]
\[
R{d\sigma \over {dt}}= |H_{0+,0+}|^2-|H_{0+,0-}|^2 = 2Re(H_dH_u^*)
\]
where the target nucleon transversity amplitudes $H_\tau$ are
\begin{equation}
H_u={1 \over {\sqrt {2}}}(H_{0+,0+}+iH_{0+,0-}), \quad  H_d={1 \over {\sqrt{2}}}(H_{0+,0+}-iH_{0+,0-})
\end{equation}
The helicity and transversity amplitudes in $\pi N \to \pi N$ are natural exchange amplitudes. Setting unnatural exchange amplitudes in (2.72) and (2.73) equal zero, we recover the relations (3.2).\\

The spin observables are not independent but satisfy condition $T^2+A^2+R^2=1$ which follows from (3.3). Using this condition and assuming that the initial state is pure with $|\vec {P}|^2=1$, we easily verify that $|Q(\vec {P})|^2=1$ and the final state is pure. We also observe from (3.2) that even when recoil nucleon polarization is not measured, the measurements of target asymmetry $T$ fully determine a single value of the transverse component of recoil nucleon polarization $Q^2$. From (3.3) we can see that the measurements of $T$ together with ${d\sigma/{dt}}$ yield a single solution for moduli of the transversity amplitudes, as expected from a unitary $S$-matrix.

\section{Unitarity conditions in $\pi^- p \to \pi^- \pi^+ n$ and similar processes.}

We wish to test unitarity in $\pi^- p \to \pi^- \pi^+ n$ using exact expressions (2.34) and (2.35) for angular components $I^j_k(\Omega,\vec {P})$ without truncation of the angular expansion. The final state is pure if and only if the recoil nucleon polarization  $|\vec {Q}(\Omega, \vec {P})|^2=1$, or equivalently, if and only if the intensities
$|\vec {I}(\Omega, \vec {P})|^2 = (I^0(\Omega, \vec {P}))^2$. We recall from (2.15) that
$I^j(\Omega, \vec {P})=I^j_u+I^j_xP_x+I^j_yP_y+I^j_zP_z$. We will work with two pure initial states with transverse target polarization $\vec {P}=(0,P_y,0)$ where $P_y=\pm 1$. With these initial pure states the final states will be pure if and only if
\begin{equation}
(I^1_u\pm I^1_y)^2 + (I^3_u\pm I^3_y)^2 =(I^0_u\pm I^0_y)^2 -(I^2_u\pm I^2_y)^2
\end{equation}
Using angular expansions (2.34) and (2.35) we can write these unitarity conditions in the form 
\begin{equation}
\Bigl ( \sum \limits_{J\lambda} \sum \limits_{J'\lambda'} (ImR^1_u \pm ImR^1_y)^{JJ'}_{\lambda \lambda'} Im(Y^J_\lambda Y^{J'*}_{\lambda'}) \Bigr )^2 +
\Bigl ( \sum \limits_{J\lambda} \sum \limits_{J'\lambda'} (ImR^3_u \pm ImR^3_y)^{JJ'}_{\lambda \lambda'} Im(Y^J_\lambda Y^{J'*}_{\lambda'}) \Bigr )^2
\end{equation}
\[
=\Bigl ( \sum \limits_{J\lambda} \sum \limits_{J'\lambda'} (ReR^0_u \pm ReR^0_y)^{JJ'}_{\lambda \lambda'} Re(Y^J_\lambda Y^{J'*}_{\lambda'}) \Bigr )^2 -
\Bigl ( \sum \limits_{J\lambda} \sum \limits_{J'\lambda'} (ReR^2_u \pm ReR^2_y)^{JJ'}_{\lambda \lambda'} Re(Y^J_\lambda Y^{J'*}_{\lambda'}) \Bigr )^2 
\]
Using the Table 1.d of Lutz and Rybicki~\cite{lutz78} reproduced in the Appendix A we find for the terms with $P_y=+1$
\begin{equation}
(ImR^1_u+ImR^1_y)^{JJ'}_{\lambda\lambda'}=-2\eta_\lambda \eta_{\lambda'}
Re(U^J_\lambda N^{J'*}_{\lambda'}-N^J_\lambda U^{J'*}_{\lambda'})
\end{equation}
\[
(ImR^3_u+ImR^3_y)^{JJ'}_{\lambda\lambda'}=+2\eta_\lambda\eta_{\lambda'}
Im(U^J_\lambda N^{J'*}_{\lambda'}+N^J_\lambda U^{J'*}_{\lambda'})
\]
\[
(ReR^0_u+ReR^0_y)^{JJ'}_{\lambda\lambda'}=+2\eta_\lambda\eta_{\lambda'}
Re(U^J_\lambda U^{J'*}_{\lambda'}+N^J_\lambda N^{J'*}_{\lambda'})
\]
\[
(ReR^2_u+ReR^2_y)^{JJ'}_{\lambda\lambda'}=-2\eta_\lambda\eta_{\lambda'}
Re(U^J_\lambda U^{J'*}_{\lambda'}-N^J_\lambda N^{J'*}_{\lambda'})
\]
where $\eta_\lambda=1$ for $\lambda=0$ and $\eta_\lambda=1/\sqrt{2}$ for $\lambda \neq 0$. The relations for terms with polarization $P_y=-1$ are the same with replacements $U^J_\lambda, N^J_\lambda \to \overline {U}^J_\lambda, \overline {N}^J_\lambda$. In the following we will work with the unitarity condition (4.2) for polarization $P_y=+1$ which involves only amplitudes with target nucleon transversity "up". The condition for the polarization $P_y=-1$ will have the same final form involving only amplitudes with target nucleon transversity "down". \\

After relabeling, the two terms on l.h.s. of (4.2) have the form
\begin{equation}
16R^2 \equiv \Bigl ( \sum \limits_{J\lambda} \sum \limits_{J' \lambda'} -2\eta_\lambda \eta_{\lambda'}(Re(U^J_\lambda N^{J'*}_{\lambda'})-Re(N^J_\lambda U^{J'*}_{\lambda'}))
Im(Y^J_\lambda Y^{J'*}_{\lambda'}) \Bigr )^2
\end{equation}
\[
=16\Bigl ( \sum \limits_{J\lambda} \sum \limits_{K \mu} \eta_\lambda \eta_{\mu} Re(U^J_\lambda N^{K*}_{\mu})
Im(Y^J_\lambda Y^{K*}_{\mu}) \Bigr )^2
\]
\begin{equation}
16I^2 \equiv \Bigl ( \sum \limits_{J\lambda} \sum \limits_{J' \lambda'} +2\eta_\lambda \eta_{\lambda'}(Im(U^J_\lambda N^{J'*}_{\lambda'})+Im(N^J_\lambda U^{J'*}_{\lambda'}))
Im(Y^J_\lambda Y^{J'*}_{\lambda'}) \Bigr )^2
\end{equation}
\[
=16\Bigl ( \sum \limits_{J\lambda} \sum \limits_{K \mu} \eta_\lambda \eta_{\mu} Im(U^J_\lambda N^{K*}_{\mu})
Im(Y^J_\lambda Y^{K*}_{\mu}) \Bigr )^2
\] 
Taking squares of the two terms on the r.h.s. of (4.2), the common summation will involve combinations
\begin{equation}
(ReR^0_u+ReR^0_y)^{JJ'}_{\lambda \lambda'}(ReR^0_u+ReR^0_y)^{KK'}_{\mu \mu'}-
(ReR^2_u+ReR^2_y)^{JJ'}_{\lambda \lambda'}(ReR^2_u+ReR^2_y)^{KK'}_{\mu \mu'}=
\end{equation}
\[
4\eta_\lambda \eta_{\lambda'} \eta_\mu \eta_{\mu'} 
\bigl (2Re(U^J_\lambda U^{J'*}_{\lambda'})Re(N^K_\mu N^{K'*}_{\mu'})+
2Re(N^J_\lambda N^{J'*}_{\lambda'})Re(U^K_\mu U^{K'*}_{\mu'})\bigr )
\]
After relabeling the second term on r.h.s. of (4.6), the r.h.s. of unitarity condition (4.2)
factorizes into a product
\begin{equation}
16UN= 16 \Bigl (\sum \limits_{J\lambda} \sum \limits_{J' \lambda'} 
Re(U^J_\lambda U^{J'*}_{\lambda'})Re(Y^J_\lambda Y^{J'*}_{\lambda'}) \Bigr )
\Bigl (\sum \limits_{K\mu} \sum \limits_{K' \mu'} 
Re(N^K_\mu N^{K'*}_{\mu'})Re(Y^K_\mu Y^{K'*}_{\mu'}) \Bigr )
\end{equation}
Using parity relations (5.52) for transversity amplitudes and the fact that $N^J_0=0$, we can rewrite the expression for the sums $R,I,U$ and $N$ in the form
\begin{equation}
R=-2\sum \limits_{J,\lambda \geq0} \sum \limits_{K,\mu >0}\xi_\lambda \eta_\lambda \eta_\mu
Re(U^J_\lambda N^{K*}_\mu)Re(Y^J_\lambda)Im(Y^K_\mu)
\end{equation}
\[
I=-2\sum \limits_{J,\lambda \geq0} \sum \limits_{K,\mu >0}\xi_\lambda \eta_\lambda \eta_\mu
Im(U^J_\lambda N^{K*}_\mu)Re(Y^J_\lambda)Im(Y^K_\mu)
\]
\[
U=\sum \limits_{J,\lambda \geq0} \sum \limits_{J',\lambda' \geq0}\xi_\lambda \xi_{\lambda'} \eta_\lambda \eta_{\lambda'}
Re(U^J_\lambda U^{J'*}_{\lambda'})Re(Y^J_\lambda)Re(Y^{J'}_{\lambda'})
\]
\[
N=4\sum \limits_{K,\mu >0} \sum \limits_{K',\mu'>0} \eta_\mu \eta_{\mu'}
Re(N^J_\mu N^{K'*}_{\mu'})Im(Y^K_\mu)Im(Y^{K'}_{\mu'})
\]
where $\xi_\lambda =1$ for $\lambda=0$ and $\xi_\lambda=2$ for $\lambda >0$. The unitarity condition $R^2+I^2=UN$ then reads
\begin{equation}
\sum \limits_{J,\lambda \geq0} \sum \limits_{J',\lambda' \geq0}\sum \limits_{K,\mu >0} \sum \limits_{K',\mu'>0}C_{\lambda \lambda', \mu \mu'}
\bigl (Re(U^J_\lambda N^{K*}_\mu)Re(U^{J'}_{\lambda'} N^{K'*}_{\mu'})+
\end{equation}
\[
Im(U^J_\lambda N^{K*}_\mu)Im(U^{J'}_{\lambda'} N^{K'*}_{\mu'})\bigr )
Y^{JJ',KK'}_{\lambda \lambda', \mu \mu'}(\Omega) 
\]
\[
=\sum \limits_{J,\lambda \geq0} \sum \limits_{J',\lambda' \geq0} \sum \limits_{K,\mu >0} \sum \limits_{K',\mu'>0}C_{\lambda \lambda', \mu \mu'}
\bigl (Re(U^J_\lambda U^{J'*}_{\lambda'})Re(N^{K*}_\mu N^{K'*}_{\mu'}) \bigr ) 
Y^{JJ',KK'}_{\lambda \lambda', \mu \mu'}(\Omega)
\]
where 
\begin{equation}
C_{\lambda \lambda', \mu \mu'}=\xi_\lambda \xi_{\lambda'} \eta_\lambda \eta_{\lambda'} \eta_\mu \eta_{\mu'}
\end{equation}
\[
Y^{JJ',KK'}_{\lambda \lambda', \mu \mu'}(\Omega) = Re(Y^J_\lambda)Re(Y^{J'}_{\lambda'})Im(Y^K_\mu)Im(Y^{K'}_{\mu'})
\]
The condition (4.9) must hold true for all solid angles $\Omega$. This is possible if and only if
\begin{equation}
Re(U^J_\lambda N^{K*}_\mu)Re(U^{J'}_{\lambda'} N^{K'*}_{\mu'})+
Im(U^J_\lambda N^{K*}_\mu)Im(U^{J'}_{\lambda'} N^{K'*}_{\mu'})=
Re(U^J_\lambda U^{J'*}_{\lambda'})Re(N^{K*}_\mu N^{K'*}_{\mu'}) 
\end{equation}
is true for all $J,J',K,K',\lambda, \lambda', \mu, \mu'$. With target polarization $P_y=-1$ we obtain from (4.2) the same unitarity condition (4.11) for transversity amplitudes 
$\overline {U}^J_\lambda$ and  $\overline {N}^K_\mu$ with target nucleon transversity "down".
Next we introduce relative phases
\begin{equation}
\alpha^{JJ'}_{\lambda \lambda'}=\Phi(U^J_\lambda)-\Phi(U^{J'}_{\lambda'}), \quad 
\beta^{KK'}_{\mu \mu'}=\Phi(N^K_\mu)-\Phi(N^{K'}_{\mu'})
\end{equation}
\[
\gamma^{JK}_{\lambda \mu}=\Phi(U^J_\lambda)-\Phi(N^K_\mu), \quad
\gamma^{J'K'}_{\lambda' \mu'}=\Phi(U^{J'}_{\lambda'})-\Phi(N^{K'}_{\mu'})
\]
and similar relative phases $\overline {\alpha}^{JJ'}_{\lambda \lambda'}$, $\overline {\beta}^{KK'}_{\mu \mu'}$, $\overline {\gamma}^{JK}_{\lambda \mu}$ and $\overline {\gamma}^{J'K'}_{\lambda' \mu'}$ for transversity amplitudes $\overline {U}^J_\lambda$ and  $\overline {N}^K_\mu$. With
\begin{equation}
\gamma^{JK}_{\lambda \mu}-\gamma^{J'K'}_{\lambda' \mu'}=
\alpha^{JJ'}_{\lambda \lambda'}-\beta^{KK'}_{\mu \mu'}, \qquad
\overline {\gamma}^{JK}_{\lambda \mu}-\overline {\gamma}^{J'K'}_{\lambda' \mu'}=
\overline {\alpha}^{JJ'}_{\lambda \lambda'}-\overline {\beta}^{KK'}_{\mu \mu'} 
\end{equation}
the two unitarity conditions (4.11) now have a final form
\begin{equation}
|U^J_\lambda||U^{J'}_{\lambda'}||N^K_\mu||N^{K'}_{\mu'}|
\cos(\alpha^{JJ'}_{\lambda \lambda'}-\beta^{KK'}_{\mu \mu'})=
|U^J_\lambda||U^{J'}_{\lambda'}||N^K_\mu||N^{K'}_{\mu'}|
\cos(\alpha^{JJ'}_{\lambda \lambda'}) \cos(\beta^{KK'}_{\mu \mu'})
\end{equation}
\[
|\overline {U}^J_\lambda||\overline {U}^{J'}_{\lambda'}||\overline {N}^K_\mu|
|\overline {N}^{K'}_{\mu'}|
\cos(\overline {\alpha}^{JJ'}_{\lambda \lambda'})-\overline {\beta}^{KK'}_{\mu \mu'})=
|\overline {U}^J_\lambda||\overline {U}^{J'}_{\lambda'}||\overline {N}^K_\mu|
|\overline {N}^{K'}_{\mu'}|
\cos(\overline {\alpha}^{JJ'}_{\lambda \lambda'}) \cos(\overline {\beta}^{KK'}_{\mu \mu'})
\]
The unitarity conditions (4.14) will be satisfied in two cases. In the case that all moduli are non-zero, the conditions (4.14) require 
\begin{equation}
\sin(\alpha^{JJ'}_{\lambda \lambda'}) \sin(\beta^{KK'}_{\mu \mu'})=0, \qquad
\sin(\overline {\alpha}^{JJ'}_{\lambda \lambda'}) \sin(\overline {\beta}^{KK'}_{\mu \mu'})=0
\end{equation}
The conditions (4.14) will also be satisfied when only one unnatural or only one natural exchange amplitude is non-zero. In such a case
\begin{equation}
|U^J_\lambda|=|U^{J_u}_{\lambda_u}|\delta_{JJ_u} \delta_{\lambda \lambda_u}, \quad 
|\overline {U}^J_\lambda|=|\overline {U}^{J_d}_{\lambda_d}|\delta_{JJ_d} \delta_{\lambda \lambda_d} \text{  or  }
\end{equation}
\[
|N^K_\mu|=|N^{K_u}_{\mu_u}|\delta_{KK_u} \delta_{\mu \mu_u}, \quad 
|\overline {N}^K_\mu|=|\overline {N}^{K_d}_{\mu_d}|\delta_{KK_d} \delta_{\mu \mu_d}
\]
Unitarity of the $S$-matrix thus requires that in $\pi^- p \to \pi^- \pi^+ n$ and other similar processes either the condition (4.15) (requirement A) or the condition (4.16) (requirement B) be satisfied for all $J,J',K,K',\lambda, \lambda', \mu, \mu'$ at all energies $s$, momentum transfers $t$ and dimeson masses $m$.

\section{Tests of unitarity conditions in measurements of $\pi^- p \to \pi^- \pi^+ n$ on polarized target.}

To discuss the experimental results we first introduce the spectroscopic notation for the amplitudes
\begin{equation}
S_\tau=U^0_{0,\tau},P^0_\tau=U^1_{0,\tau}, P^U_\tau=U^1_{1,\tau},P^N_\tau=N^1_{1,\tau}
\end{equation}
\[
D^0_\tau=U^2_{0,\tau},D^U_\tau=U^2_{1,\tau},D^N_\tau=N^2_{1,\tau},
D^{2U}_\tau=U^2_{2,\tau},D^{2N}_\tau=N^2_{2,\tau}
\]
where $\tau$ is the target transversity $\tau=u,d$. The CERN measurements and amplitude analysis of $\pi^- p \to \pi^- \pi^+ n$ on polarized target at 17.2 GeV/c fall into 3 kinematic regions.\\

In the first region dipion masses are in the range of 580-900 MeV at low $|t|$ ($0.005 \leq |t| \leq 0.2$  (GeV/c$)^2$) and in the $|t|$ range of $0.0 \leq |t| \leq 1.0$ (GeV/c$)^2$ at $\rho^0(770)$ mass. In both these subregions $S$- and $P$-wave dominate. Amplitude analyses of this data~\cite{becker79a,svec96,svec97a,svec02a} all agree on non-zero moduli for all amplitudes and non-zero relative phases between all unnatural exchange amplitudes. Since there is only one natural this data do not violate unitarity conditions (4.15) and (4.16). Explicit calculation shows that $|Q(\Omega, \vec{P})|^2=1$ for pure states $\vec {P}=(0,\pm 1,0)$ for all solutions of amplitudes. Similar results were obtained in the amplitude analyses of CERN measurements of 
$\pi^+n \to \pi^+ \pi^- p$~\cite{lesquen85,svec92a,svec96,svec97a} at 5.98 and 11.85 GeV/c, and
$K^+ n \to K^+ \pi^- p$~\cite{lesquen89,svec92b} at 5.98 GeV/c for dimeson masses below 1000 MeV.\\

In the second region the dipion masses extend from 580 - 1780 MeV at low momentum transfers 
$0.01 \leq |t| \leq 0.20$ (GeV/c$)^2$. Two amplitude analyses of the data with different mass binnings~\cite{becker79b,chabaud83} found non-zero moduli of all amplitudes with helicities less than 2, and reported non-zero relative phases between unnatural echange amplitudes. There are two pairs of natural exchange amplitudes $P^N_\tau$ and $D^N_\tau$, $\tau=u,d$ whose relative phases $\Phi (P^N)-\Phi (D^N)$ are not reported. Unitarity condition (4.15) requires that the phases $\beta^{12}_{11}=\Phi (P^N_u)-\Phi (D^N_u)$ and 
$\overline {\beta}^{12}_{11}=\Phi (P^N_d)-\Phi (D^N_d)$ vanish. The data violate the unitarity condition (5.16) but in the absence of data on $\Phi (P^N)-\Phi (D^N)$ the unitarity condition (4.15) cannot be tested in this kinematic region.\\

In the third region the dipion masses are in the range 580-1480 MeV at large momentum transfers $0.2\leq |t| \leq 1.0$ (GeV/c$)^2$. The analysis of Rybicki and Sakrejda of this data~\cite{rybicki85} found non-zero moduli of all amplitudes including helicity 2 amplitudes $D^{2U}_\tau$ and $D^{2N}_\tau$, a finding which violates unitarity condition (4.16). In addition to non-zero relative phases between unnatural amplitudes they report a non-vanishing phases $\Phi(P^N_u)-\Phi(D^N_u)$ and $\Phi(P^N_d)-\Phi(D^N_d)$, a finding that violates the unitarity condition (4.15). On the basis of this evidence we conclude that pure states can evolve into mixed states in $\pi^- p \to \pi^- \pi^+ n$ and that unitarity is violated.\\

\begin{figure}
\includegraphics[width=12cm,height=10.5cm]{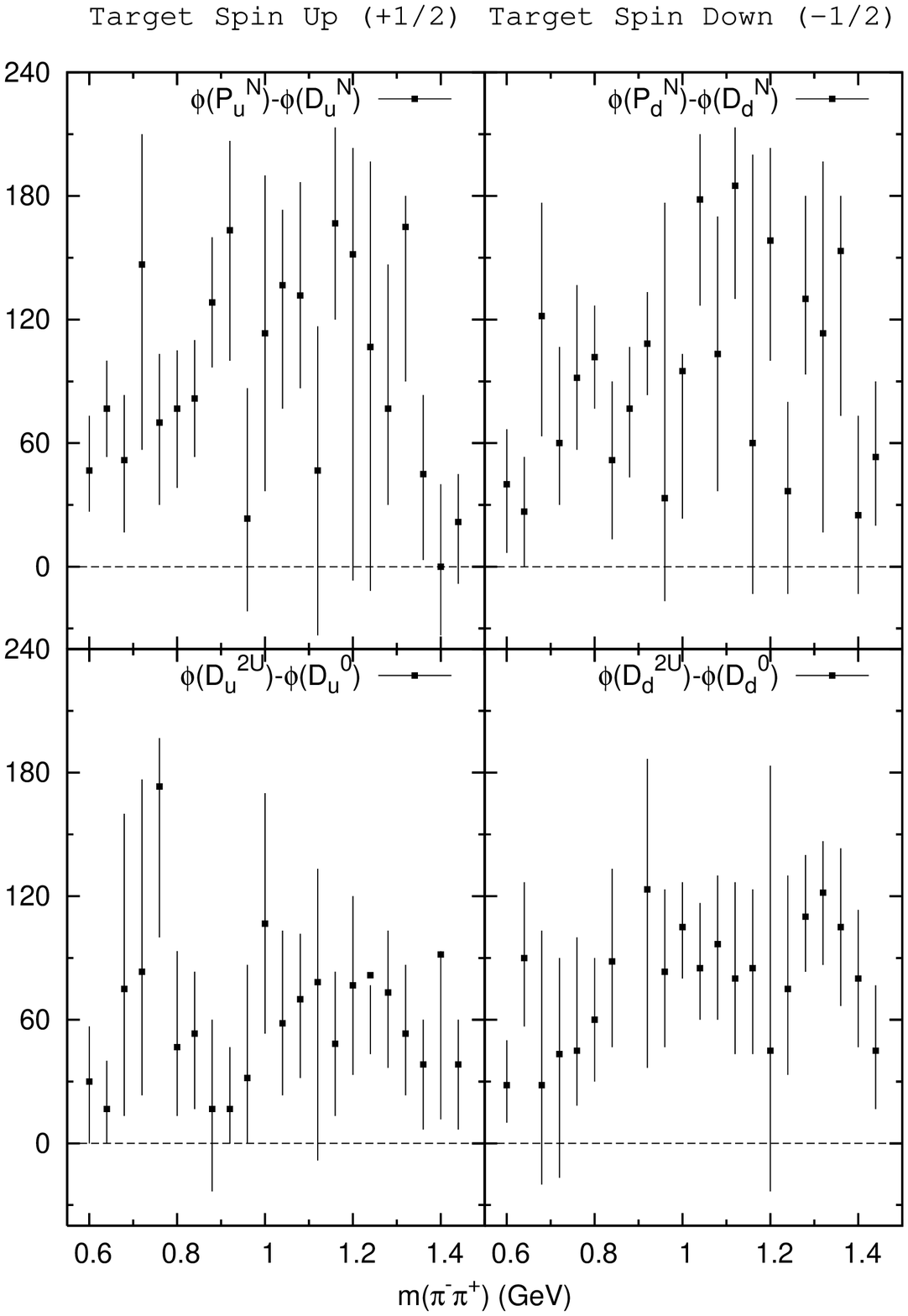}
\caption{Relative phases $\Phi (P^N_\tau)-\Phi (D^N_\tau)$ and $\Phi (D^{2U}_\tau)-\Phi (D^0_\tau)$. Data from Rybicki and Sakrejda~\cite{rybicki85}.}
\label{Figure 1.}
\end{figure}

\begin{figure}
\includegraphics[width=12cm,height=10.5cm]{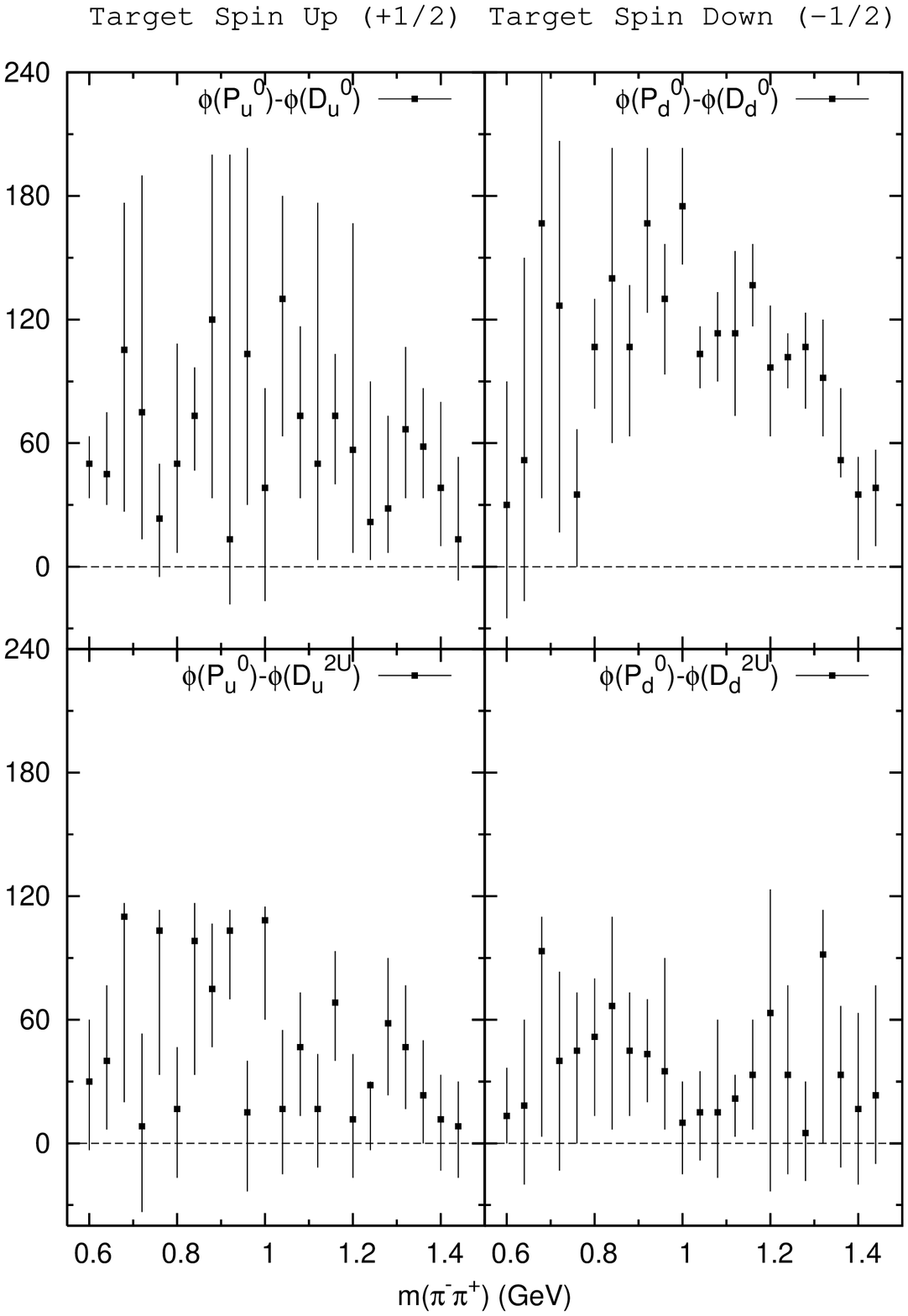}
\caption{Relative phases $\Phi (P^0_\tau)-\Phi (D^0_\tau)$ and $\Phi (P^0_\tau)-\Phi (D^{2U}_\tau)$. Data from Rybicki and Sakrejda~\cite{rybicki85}.}
\label{Figure 2.}
\end{figure}

Figures 1 and 2 show selected relative phases from the analysis of Rybicki and Sakrejda. The phases $\Phi(P^N_u)-\Phi(D^N_u)$ and $\Phi(P^N_d)-\Phi(D^N_d)$ are shown in Figure 1. Phases $\Phi(D^{2U}_\tau)-\Phi(D^0_\tau)$,$\Phi(P^0_\tau)-\Phi(D^0_\tau)$ and $\Phi(P^0_\tau)-\Phi(D^{2U}_\tau)$ shown in Figures 1 and 2 illustrate the relative phases between unnutural exchage amplitudes. They will be used in Section VIII. to confirm a crucial prediction arising from the principal consequence of violation of unitarity, namely that pion creation processes behave as open quantum systems interacting with a quantum environment.\\

We close this discussion with a comment on BNL measurement of $\pi^- p \to \pi^0 \pi^0 n$ at 18.3 GeV/c on unpolarized target~\cite{gunter01}. In this high statistics experiment the dipion mass is in the range 350-2200 MeV for several binnings of momentum transfers at low and high $|t|$. The analysis is made on assumption that amplitudes with dipion helicity 2 do not contribute in $D$-wave and amplitudes with nonzero dipion helicity do not contribute in $G$-wave. When this assumption is removed the analysis yields non-zero amplitudes in all waves, albeit with ambiguities that cannot be resolved by the measurements on unpolarized target. Measurements of this process on polarized target~\cite{svec97d} may resolve these ambiguities and thus provide independent evidence for violation of unitarity in pion creation processes.

\section{Unitarity, superscattering operators and the metric.}

With the spinor representations (2.9) and (2.13) of the target and recoil nucleon spin density matrices $\rho_i(\vec {P})=\rho_p(\vec {P})$ and $\rho_f(\vec {P})= \rho_n(\vec {Q}(\vec {P}))I^0(\vec {P})$ we can associate four-vectors $P^\mu=(1,\vec {P})$ and $Q^\alpha I^0=(1,\vec{Q})I^0$, respectively, representing the polarization four-vectors in the rest frame of the particles. The determinants of the density matrices are the Lorentz invariant norm of the four-vectors
\begin{equation}
\det (\rho_p(\vec {P}))=1-|\vec{P}|^2=\eta_{\mu \nu}P^\mu P^\nu=\|P\| ^2
\end{equation}
\[
\det (\rho_n(\vec {Q}))(I^0)^2=(1-|\vec{Q}|^2)(I^0)^2=\eta_{\alpha \beta}Q^\alpha Q^\beta (I^0)^2=\|Q\|^2 (I^0)^2
\]
where $\eta_{\mu \nu} = diag (+1,-1,-1,-1)$ is metric in Minkovski space-time. The target and recoil nucleon polarization four-vectors are connected by the equations (2.15)
\begin{equation}
Q^\alpha (\Omega, \vec {P})I^0(\Omega, \vec {P})= I^\alpha_\mu (\Omega) P^\mu
\end{equation}
In matrix form we can write  $QI^0=\mathbf {I} P$ so that the matrix $\mathbf {I}$ has the meaning of Hawking's superscattering operator connecting initial and final states~\cite{hawking82}. The norm $\|Q\|^2 (I^0)^2$ now takes the form 
\begin{equation}
\|Q\|^2 (I^0)^2=\eta_{\alpha \beta} I^\alpha_\mu I^\beta_\nu P^\mu P^\nu = 
G_{\mu \nu}(\Omega) P^\mu P^\nu
\end{equation}
Unitary $S$-matrix requires that any pure state $\|P\|^2=0$ evolves into a pure state $\|Q\|^2=0$. A sufficient condition for this to happen is for the correlation matrix 
\begin{equation}
G_{\mu \nu} (\Omega)=Z(\Omega) \eta_{\mu \nu}
\end{equation}
for then $\|Q\|^2 (I^0)^2=Z(\Omega) \|P\|^2=0$. To find out what could be the matrix $I^\alpha_\mu$ that leaves invariant the Minkovski metric in (6.4) we turn to $\pi N \to \pi N$. The transformation (6.2) now has the form (3.2) with the matrix $I^\alpha_\mu$ given by
\begin{equation}
\mathbf {I}={d\sigma \over {dt}}
\left( \begin{array}{cccc}
1 & 0 & T & 0 \\
0 & R & 0 & -A \\
T & 0 & 1 & 0  \\
0 & A & 0 & R
\end{array} \right)
\end{equation}
The matrix (6.5) is orthogonal but not unitary: while $\sum_\mu I^\alpha _\mu I^\beta_ \mu =0$ for $\alpha \neq \beta$, $\sum_\mu I^0 _\mu I^0_\mu = \sum_\mu I^2 _\mu I^2_\mu = (1+T^2)({d\sigma /{dt}})^2$ and $\sum_\mu I^1 _\mu I^1_\mu = \sum_\mu I^3 _\mu I^3 _\mu = (1-T^2)({d\sigma /{dt}})^2$ with determinant equal to $(1-T^2)^2 ({d\sigma/ {dt}})^4$. We easily verify that the matrix (6.5) preserves the Minkovski metric
$\eta_{\alpha \beta} I^\alpha _\mu I^\beta_\nu = (1-T^2)({d\sigma /{dt}})^2\eta_{\mu \nu}$.\\

If we impose the structure of matrix (6.5) on the matix $I^\alpha_\mu (\Omega)$ in (6.2),
we obtain constraints on its elements which imply constraints on the amplitudes. From the requirements that $I^1_u=I^1_y=0$ and $I^3_u=I^3_y=0$ we obtain, using the Table 1.d of Lutz-Rybicki in Appendix A, conditions
\begin{equation}
Re(U^J_\lambda N^{J'*}_{\lambda'}-N^J_\lambda U^{J'*}_{\lambda'} -
\overline {U}^J_\lambda \overline {N}^{J'*}_{\lambda'}+\overline {N}^J_\lambda \overline {U}^{J'*}_{\lambda'})=
Re(U^J_\lambda N^{J'*}_{\lambda'}-N^J_\lambda U^{J'*}_{\lambda'} +
\overline {U}^J_\lambda \overline {N}^{J'*}_{\lambda'}-\overline {N}^J_\lambda \overline {U}^{J'*}_{\lambda'})= 0
\end{equation}
\[
Im(U^J_\lambda N^{J'*}_{\lambda'}+N^J_\lambda U^{J'*}_{\lambda'} +
\overline {U}^J_\lambda \overline {N}^{J'*}_{\lambda'}+\overline {N}^J_\lambda \overline {U}^{J'*}_{\lambda'})=
Im(U^J_\lambda N^{J'*}_{\lambda'}+N^J_\lambda U^{J'*}_{\lambda'} -
\overline {U}^J_\lambda \overline {N}^{J'*}_{\lambda'}-\overline {N}^J_\lambda \overline {U}^{J'*}_{\lambda'})= 0
\]
From these conditions we find
\begin{equation}
Re(N^J_\lambda U^{J'*}_0)=Im(N^J_\lambda U^{J'*}_0)=0, \qquad
Re(\overline {N}^J_\lambda \overline {U}^{J'*}_0)=Im(\overline {N}^J_\lambda \overline {U}^{J'*}_0)=0
\end{equation}
These conditions require that either all unnatural exchange ampliyudes with zero helicity or all natural exchange amplitudes vanish, in a clear contradiction with known data discussed in the previous Section.\\

Based on the violation of unitarity conditions for the special transverse initial polarization $P_y= \pm1$, we conjecture that in pion creation processes the superscattering operatotors in (6.2) do not preserve Minkovski metric and thus evolve any initial pure state into a mixed final state. Formally, the particle intraction results in a non-unitary transformation of the four-vector $P_\mu$ into the four-vector $Q^\alpha$ which transforms Minkovski metric $\eta_{\alpha \beta}$ into metric $G_{\mu \nu}(\Omega)$ in a curved space-time associated with the interaction process.

\section{Non-unitary evolution of open quantum systems.}

In this Section we briefly review how evolution from pure states into mixed states occurs in the theory of open quantum systems interacting with a quantum environment. These concepts are essential for the physical interpretation of the violations of unitarity conditions and the reinterpretation of the spin formalism developed in the Section II. to form a new picture of pion creation processes as open quantum systems interacting with a quantum environment. The discussion follows largely the book by Nielsen and Chuang~\cite{nielsen00}.\\

Associated with any Hilbert space $H$ of state vectors with basis states $|m>$ is a Hilbert-Schmidt space ${\cal B}(H)$ of density matrix operators with basis operators $|m><n|$ and inner product $(A,B)=Tr(A^+B)$~\cite{fano57,blum96,nielsen00}. Density matrix operator $\rho$ represents the quantum state of any quantum system. It is a positive definite hermitian matrix that satisfies a condition~\cite{leader01,fano57,blum96,nielsen00}
\begin{equation}
Tr(\rho^2) \leq (Tr(\rho))^2
\end{equation}
A state about which we can have a complete knowledge is said to be a pure state.  A state of which we can have only an incomplete knowledge is called mixed state. The equality $Tr(\rho^2)=(Tr(\rho))^2$ is satisfied if and only if the state is pure. A quantum state is represented also by a state vector $|\Psi>$ in a Hilbert space $H$ if and only if the state $\rho$ is  pure. In that case $\rho=|\Psi><\Psi|$. The Hilbert-Schmidt space ${\cal B}(H)$ includes mixed states to which correspond no vector states in $H$.\\

Unitary $S$-matrix connects initial vector states $|\psi_{in}>$ in $H_{in}$ to final vector states $|\psi_{out}>$ in $H_{out}$ and the corresponding pure states $\rho_{in}$ and $\rho_{out}$ in ${\cal B}(H_{in})$ and ${\cal B}(H_{out})$, respectively. It also connects mixed initial states $\rho_{in}$ with mixed final states $\rho_{out}$ since such states are incoherent sum of corresponding pure states. The question arises whether a linear mapping from ${\cal B}(H_{in})$ to ${\cal B}(H_{out})$ exists that preserves superposition principle but is more general than the $S$-matrix, and what physical reality it would represent. Such mappings in fact exist and describe generally non-unitary evolution  $\rho_{out}(S)={\cal E}(\rho_{in}(S))$ of an open quantum system or process $S$ interacting with an environment 
$E$~\cite{nielsen00,kraus83,breuer02}.\\

The co-evolution of an open quantum system $S$ and a quantum environment $E$ is assumed to be described by a unitary operator $U$. The initial state $\rho_i(S,E)$ of the combined system is prepared in a separable state $\rho_i(S,E)=\rho_i(S) \otimes \rho_i(E)$.  A unitary quantum operation ${\cal E}$ describing the interaction of $S$ and $E$ evolves $\rho_i(S,E)$ into a final state~\cite{nielsen00,kraus83,breuer02}
\begin{equation}
\rho_f(S,E)= {\cal E}(\rho_i(S,E))=U \rho_i(S,E) U^+ 
\end{equation}
\noindent
$\rho_f(S,E)$ is not a separable but an entangled state of $S$ and $E$. The observer can perform measurements on the system $S$ but cannot perform direct measurements on the environment $E$. After the transformation $U$, the system $S$ no longer interacts with environment $E$. The quantum state of system $S$ is then fully described by reduced density matrix
\begin{equation}
\rho_f(S)=Tr_E (\rho_f(S,E)) = {\cal E} (\rho_i(S))
\end{equation}
\noindent
in a sense that we can calculate average values $<Q>=Tr(Q \rho_f(S))$ of any observable $Q$. The trace in (7.3) is over the interacting degrees of freedom $|e_\ell>$ of the environment which form an orthonormal basis in the finite dimensional Hilbert space $H(E)$ of the environment $E$. In general, the reduced state $\rho_f(S)$ is no longer related to the initial state $\rho_i(S)$ by a unitary transformation $\rho_f=S\rho_i S^+$. Instead it is given by the non-unitary Kraus representation~\cite{nielsen00,kraus83,breuer02} that generalizes the evolution equation $\rho_f=S\rho_i S^+$
\begin{equation}
\rho_f(S)=\sum \limits_{\ell} S_\ell \rho_i(S) S_\ell^+
\end{equation}
\noindent
where $S_\ell=<e_\ell |U|e_0>$ are operators acting on the state space of the system $S$ and where we assumed that $\rho_i(E)=|e_0><e_0|$ is a pure state. Physically Kraus representation corresponds to a measurement of the environment $E$ by the system $S$ just after the unitary co-evolution $U$. The operator elements ${S_\ell}$ must satisfy the completness relation
\begin{equation}
\sum \limits_{\ell} S_\ell^+ S_\ell = I
\end{equation}
for trace preserving quantum operations ${\cal E} (\rho_i(S,E))$. When the initial state of the environment is a mixed state $\rho_i(E)=\sum \limits_{m,n} p_{mn}|e_m><e_n|$, the Kraus representation takes the form~\cite{nielsen00}
\begin{equation}
\rho_f(S)=\sum \limits_{\ell} \sum \limits_{m,n} p_{mn} S_{\ell m} \rho_i(S) S^+_{n\ell}
\end{equation}
where the operators $S_{\ell m}=<e_\ell |U|e_m>$ satisfy a completness relation similar to (7.5) and
\begin{equation}
Tr(\rho_i(E))=\sum\limits_m p_{mm} = 1
\end{equation}
When certain degrees of freedom $S^{''}$ of the system $S$ are not measured, the density matrix is further reduced by taking a trace over $S^{''}$. In certain cases the resulting reduced density matrix $\rho_f(S^{'})$ has a part in the unitary form $\rho_f=S\rho_i S^+$ independent of the interacting degrees of freedom of the environment. This part is referred to as a decoherence free subspace of Hilbert-Schmidt space ${\cal B}(H)$.\\

The Hilbert space $H(E)$ of the environment $E$ is formed by the interacting degrees of freedom $|e_\ell>$ involved in its interaction with the system or process $S$. The Hilbert space $H(E)$ has a finite dimension $\dim H(E) \leq \dim H_i(S) 
\dim H_f(S)$~\cite{nielsen00}. Interactions of $S$ with $E$ in which there is an exchange of energy and/or momentum are called dissipative interactions. In contrast, there is no exchange of energy or momentun in non-dissipative interactions of $S$ with $E$ . Non-dissipative interactions are also referred to as dephasing interactions since in general they effect only a change of phases.

\section{Pion creation processes as open quantum systems interacting with a quantum environment.}

Assuming unitary $S$-matrix and that the pion creation process $\pi^- p \to \pi^- \pi^+ n$  behaves as an isolated quantum system $S$, we used evolution equation
\begin{equation}
\rho_f(S)=S\rho_i(S) S^+
\end{equation}
to develop in Section II. a spin formalism to express $\rho_f(S) \equiv \rho_f(\Omega, \vec {P})$ in terms of measurable spin density matrix elements $(R^j_k)^{JJ'}_{\lambda \lambda'}$ which are bilinear combinations of transversity amplitudes with definite $t$-channel naturality $U^J_{\lambda, \tau}$ and $N^J_{\lambda, \tau}$. In developing this formalism we used the form (8.1) but we made no explicit use of the unitarity of the $S$-matrix or of the assuption that the pion creation process is an isolated event.\\

Unitary $S$-matrix evolves isolated system in any pure state into an isolated system in a pure state. In Section IV. we imposed this condition on $\rho_f(\Omega, \vec {P})$ for specific initial pure states with nucleon polarization $\vec {P}=(0,\pm1,0)$ and obtained unitarity conditions (4.15) and (4.16) that are violated by the CERN data on polarized target. We concluded that pure inital states $\rho_i$ evolve into mixed final states $\rho_f$ in pion creation processes. This means the evolution in these processes is not unitary. In quantum theory such non-unitary evolution occurs in open quantum systems interacting with an environment. We are thus led to the conclusion that pion creation processes are behaving as open quantum systems $S$ interacting with a quantum environment $E$ in a unitary co-evolution
described by a co-evolution operator $U$. The observed final state $\rho_f(S)$ is then reduced density matrix given by Kraus representation (7.6)
\begin{equation}
\rho_f(S)=Tr_E(\rho_f(S,E))=\sum \limits_{\ell} \sum \limits_{m,n} p_{mn} S_{\ell m} \rho_i(S) S^+_{n\ell}=\sum \limits_\ell \sum \limits_{m.n} p_{mn}\rho_f(\ell m,n\ell)
\end{equation}

To summarize, on one hand we have the spin formalism developed in Section II. from the unitary evolution equation (8.1) that has been used in amplitude analyses of the data. On the other hand, the data analyses lead to the conclusion that $\rho_f$ is given by a non-unitary Kraus representation (8.2). Self-consistency requires that Kraus representation leaves invariant the experimental form of angular intensities given by (2.40) and (2.41) and the form of equations for density matrix elements in terms of amplitudes given by the Lutz-Rybicki tables. In this Section we show that such invariance holds provided that the unitary co-evolution conserves $P$-parity and quantum numbers of the environment.\\

Each term in Kraus representation (8.2) of the mixed state $\rho_f(S)$ has the formal form of the evolution equation (8.1). Thus we can apply the spin formalism of the Section II. to each term $\rho_f(\ell m,n\ell)=S_{\ell m} \rho_i(S) S^+_{n\ell}$ separately with separate co-evolution helicity amplitudes
\begin{equation}
H^J_{\lambda \chi,0 \nu}(\ell m)=<J \lambda, \chi|S_{\ell m}|0 \nu>=
<J \lambda, \chi|<e_\ell|U|e_m>|0 \nu>
\end{equation}
and the corresponding co-evolution transversity amplitudes $U^J_{\lambda, \tau}(\ell m)$ and $N^J_{\lambda, \tau}(\ell m)$. Each term in (8.2) has the form (2.11)
\begin{equation}
\rho_f(\theta \phi, \vec{P},\ell m,n\ell) = {1 \over{2}} \bigl (I^0 (\theta \phi, \vec{P})_{\ell m,n\ell} \sigma^0 + {\vec{I}} (\theta \phi, \vec{P})_{\ell m,n\ell} \vec{\sigma}\bigr )
\end{equation}
with a decomposition (2.15) for the intensities
\begin{equation}
I^j(\theta \phi, \vec{P})_{\ell m,n\ell} = Tr(\sigma^j \rho_f(\theta \phi, \vec{P},\ell m,n\ell))
\end{equation}
\[
=I^j_u(\theta \phi)_{\ell m,n\ell} + P_x I^j_x(\theta \phi)_{\ell m,n\ell}+ P_y I^j_y(\theta \phi)_{\ell m,n\ell}+ P_z I^j_z(\theta \phi)_{\ell m,n\ell}
\]
The polarization components of intensities $I^j_k(\theta \phi)_{\ell m,n\ell}$ have angular expansion similar to (2.24)
\begin{equation}
I^j_k(\theta \phi)_{\ell m,n\ell}=\sum \limits_{J \lambda} \sum \limits_{J' \lambda'} (R^j_k(\ell m,n\ell))^{JJ'}_{\lambda \lambda'} Y^J_{\lambda}(\theta \phi) Y^{J'*}_{\lambda'}(\theta \phi)
\end{equation}
where the unnormalized angular density matrix elements $R^j_k(\ell m,n\ell)$ are expressed in terms of co-evolution helicity amplitudes by relations similar to (2.26)
\begin{equation}
(R^j_k(\ell m,n\ell))^{JJ'}_{\lambda \lambda'}= {1 \over{2}} \sum \limits_{\chi,\chi''} \sum \limits_{\nu \nu'} (\sigma^j)_{\chi\chi''}H^J_{\lambda \chi'', 0 \nu}(\ell m)(\sigma_k)_{\nu \nu'} H^{J'*}_{\lambda' \chi, 0 \nu'}(n\ell)
\end{equation}
As the result of linearity of the Kraus representation, the observed final state density matrix still has the form (2.11)
\begin{equation}
\rho_f(\theta \phi, \vec{P}) = {1 \over{2}} \bigl (I^0 (\theta \phi, \vec{P}) \sigma^0 + {\vec{I}} (\theta \phi, \vec{P}) \vec{\sigma}\bigr )
\end{equation}
where
\begin{equation}
I^j_k(\theta \phi,\vec {P})=\sum \limits_\ell \sum \limits_{mn} p_{mn}
(I^j_k(\theta \phi))_{\ell m,n\ell}
\end{equation}
The observed intensities $I^j_k(\theta \phi, \vec {P})$ are co-evolution intensities $I^j_k(\theta \phi, \vec {P})_{\ell m,n\ell}$ averaged over the initial state of the environment and summed over its final states. There are four interacting degrees of freedom of the environment $|e_\ell>, \ell =1,4$ allowed by the condition $\dim H(E) \leq \dim H_i(S)\dim H_f(S)=(2s_p+1)(2s_n+1)=4$.\\

To bring $I^j_k(\theta \phi, \vec {P})$ to the form (2.40) and (2.41), we need to bring to this form the co-evolution intensities $I^j_k(\theta \phi, \vec {P})_{\ell m,n\ell}$. To this end we follow the proceedure of Section II.C. First we arrange the sum (8.6) in the form (2.28). To bring this form to the form (2.30) we need a hermiticity relation
\begin{equation}
(R^j_k(\ell m, n\ell))^{J'J}_{\lambda' \lambda}=(R^j_k(\ell m, n\ell))^{JJ'*}_{\lambda \lambda'}
\end{equation}
The actual calculation using (8.7) gives
\begin{equation}
(R^j_k(\ell m, n\ell))^{J'J}_{\lambda' \lambda}=(R^j_k(n\ell,\ell m))^{JJ'*}_{\lambda \lambda'}
\end{equation}
The hermiticity condition (8.8) can be satisfied only when the amplitudes have a diagonal form
\begin{equation}
H^J_{\lambda \chi,0\nu}(\ell m)=H^J_{\lambda \chi,0\nu}(\ell \ell)\delta_{\ell m}\equiv
H^J_{\lambda \chi,0\nu}(\ell)\delta_{\ell m}
\end{equation}
Physically this means that the quantum numbers of the quantum states $|e_\ell>, \ell=1,4$  must be conserved in the unitary co-evolution process $U$.\\ 

To proceed next, the co-evolution density matrix elements $R^j_k(\ell,\ell))^{JJ'}_{\lambda \lambda'}$ must satisfy parity relations (2.32) and (2.33). This requires that $P$-parity is conserved in the co-evolution process so that the co-evolution amplitudes satisfy parity relations similar to (2.31)
\begin{equation}
H^J_{-\lambda -\chi, 0 -\nu}(\ell)=(-1)^{\lambda+\chi+\nu}H^J_{\lambda \chi, 0 \nu}(\ell)
\end{equation}
With diagonal form of amplitudes (8.12) the components of intensities $I^j_k(\theta \phi)_{\ell m,n\ell}$ have a diagonal form
\begin{equation}
I^j_k(\theta \phi)_{\ell m,n\ell}=I^j_k(\theta \phi)_{\ell ,\ell}\delta_{\ell m}\delta_{n \ell}
\end{equation}
The diagonal components $I^j_k(\theta \phi)_{\ell ,\ell}$ now have the desired form (2.40) 
\begin{equation}
I^j_k(\theta \phi)_{\ell ,\ell} = 
\sum \limits_{J \leq J'} \sum \limits _{\lambda \geq 0} \sum \limits_{\lambda'} 
\xi_{JJ'} \xi_\lambda (Re R^j_k(\ell ,\ell))^{JJ}_{\lambda \lambda'} Re(Y^J_\lambda(\theta\phi)Y^{J*}_{\lambda'}(\theta \phi))
\end{equation}
for $(k,j)=(u,0),(y,0),(u,2),(y,2),(x,1),(z,1),(x,3),(z,3)$ and the form (2.41)
\begin{equation}
I^j_k(\theta \phi)_{\ell ,\ell} = 
\sum \limits_{ J < J'} \sum \limits_{\lambda \geq 0} \sum \limits_{\lambda'} 
\xi_{JJ'} \xi_\lambda (Im R^j_k(\ell ,\ell))^{JJ'}_{\lambda \lambda'} 
Im(Y^J_\lambda(\theta \phi)Y^{J'*}_{\lambda'}(\theta \phi))
\end{equation}
for $(x,0),(z,0),(x,2),(z,2),(u,1),(y,1),(u,3),(y,3)$. The measured intensities $I^j_k(\theta \phi)$ will thus have the same angular expansions (2.40) 
\begin{equation}
I^j_k(\theta \phi) = 
\sum \limits_{J \leq J'} \sum \limits _{\lambda \geq 0} \sum \limits_{\lambda'} 
\xi_{JJ'} \xi_\lambda (Re R^j_k)^{JJ}_{\lambda \lambda'} Re(Y^J_\lambda(\theta\phi)Y^{J*}_{\lambda'}(\theta \phi))
\end{equation}
for $(k,j)=(u,0),(y,0),(u,2),(y,2),(x,1),(z,1),(x,3),(z,3)$ and (2.41)
\begin{equation}
I^j_k(\theta \phi) = 
\sum \limits_{ J < J'} \sum \limits_{\lambda \geq 0} \sum \limits_{\lambda'} 
\xi_{JJ'} \xi_\lambda (Im R^j_k)^{JJ'}_{\lambda \lambda'} 
Im(Y^J_\lambda(\theta \phi)Y^{J'*}_{\lambda'}(\theta \phi))
\end{equation}
for $(x,0),(z,0),(x,2),(z,2),(u,1),(y,1),(u,3),(y,3)$. The measured density matrix elements $(R^j_k)^{JJ}_{\lambda \lambda'}$ in (8.17) and (8.18) are environment-averaged co-evolution density matrix elements $(R^j_k (\ell ,\ell))^{JJ'}_{\lambda \lambda'}$ 
\begin{equation}
(R^j_k)^{JJ}_{\lambda \lambda'}=\sum \limits_{\ell=1}^4  p_{\ell \ell}
(R^j_k (\ell ,\ell))^{JJ'}_{\lambda \lambda'}
\end{equation}

The co-evolution density matrix elements $(R^j_k (\ell ,\ell))^{JJ'}_{\lambda \lambda'}$ are sums of bilinear terms of co-evolution amplitudes 
$A^{J,\eta}_{\lambda,\tau} (\ell)B^{J',\eta'*}_{\lambda',\tau}(\ell)$ given in the Lutz-Rybicki tables in Appendix A. The co-evolution transversity amplitudes
\begin{equation}
A^{J,\eta}_{\lambda, \tau}(\ell)=<J\lambda \eta,\tau_n|<e_l|U|e_l>|0\tau>
\end{equation}
\[
={1 \over{\sqrt {2}}}(<J\lambda,\tau_n|<e_l|U|e_l>|0 \tau>-\eta (-1)^\lambda <J-\lambda,\tau_n|<e_l|U|e_l>|0\tau>)
\]
describe the interaction with the environment. Recall that recoil transversity $\tau_n$ is uniquely determined by target transversity $\tau$ and naturality $\eta=\pm1$ as the result of parity conservation. In the notation of Section II
\begin{equation}
U^J_{\lambda,\tau}(\ell)=A^{J,-1}_{\lambda, \tau}(\ell), \qquad
N^J_{\lambda,\tau}(\ell)=A^{J,+1}_{\lambda, \tau}(\ell)
\end{equation}

The measured density matrix elements $(R^j_k)^{JJ}_{\lambda \lambda'}$ given by (8.19) will keep the form given by Lutz-Rybicki table 1.d with moduli and bilinear terms of transversity amplitudes $A^{J,\eta}_{\lambda, \tau}$ now redefined as environment-averaged moduli and bilinear terms of co-evolution amplitudes, respectively. The equations for the measured moduli of transversity amplitudes then read
\begin{equation}
|A^{J,\eta}_{\lambda, \tau}|^2=\sum \limits_{\ell=1}^4 p_{\ell \ell}
|A^{J,\eta}_{\lambda, \tau}(\ell)|^2, \quad \tau =u,d
\end{equation}
The measured bilinear terms have the form
\begin{equation}
Re(A^{J,\eta}_{\lambda, \tau}B^{J',\eta' *}_{\lambda', \tau'})\equiv
|A^{J,\eta}_{\lambda, \tau}||B^{J',\eta' *}_{\lambda', \tau'}|
\cos(\Phi(A^{J,\eta}_{\lambda, \tau}B^{J',\eta' *}_{\lambda', \tau'}))=
\sum \limits_{\ell=1}^4  p_{\ell \ell}
Re(A^{J,\eta}_{\lambda, \tau}(\ell)B^{J',\eta'*}_{\lambda', \tau'}(\ell))
\end{equation}
\[
Im(A^{J,\eta}_{\lambda, \tau}B^{J',\eta' *}_{\lambda', \tau'})\equiv
|A^{J,\eta}_{\lambda, \tau}||B^{J',\eta' *}_{\lambda', \tau'}|
\sin (\Psi(A^{J,\eta}_{\lambda, \tau}B^{J',\eta' *}_{\lambda', \tau'}))=
\sum \limits_{\ell=1}^4 p_{\ell \ell}
Im(A^{J,\eta}_{\lambda, \tau}(\ell)B^{J',\eta'*}_{\lambda', \tau'}(\ell))
\]
The observed final state density matrix $\rho_f$ thus has the form identical to the form developed in Section II. from the evolution equation (8.1). However, it is now the averaged  density matrix elements (8.19) and the averaged moduli (8.22) and averaged correlations (8.23) that are measured in experiments on polarized targets and on which we may no longer impose unitarity conditions.

\section{Experimental test of Kraus represetation for observed final state in $\pi^- p  \to \pi^- \pi^+ n$.}

We have arrived at a crucial prediction arising from the principal consequence of violation of unitarity, namely that pion creation process behaves as an open quantum system interacting with an environment with final states given by Kraus representation. The prediction that the measured bilinear terms are environment-averaged bilinear terms of co-evolution amplitudes is testable using already existing data on $\pi^- p \to \pi^- \pi^+ n$ at large dipion masses.\\

\begin{figure}
\includegraphics[width=12cm,height=10.5cm]{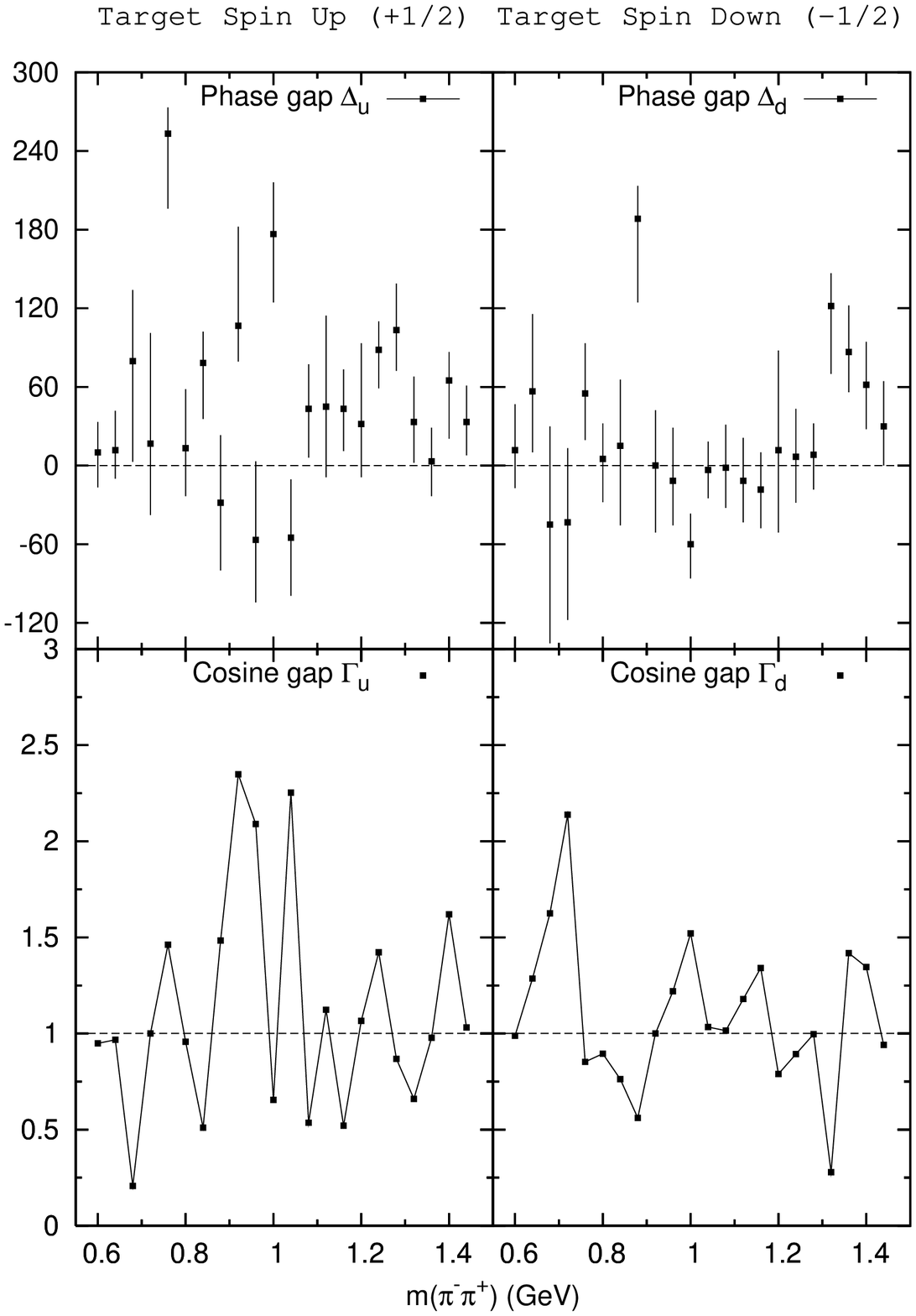}
\caption{Phase gap $\Delta_\tau$ and cosine gap $\Gamma_\tau$ for aplitudes $P^0_\tau$, $D^0_\tau$ and $D^{2U}_\tau$ at large momentum transfers $0.2 \leq |t| \leq 1.0$ (GeV/c$)^2$. Data from Rybicki and Sakrejda~\cite{rybicki85}.}
\label{Figure 3.}
\end{figure}

The CERN measurements on transversely polarized targets measure correlations of the type $Re(AB^*)=|A||B|\cos(\Phi(AB^*))$ where $\cos(\Phi(AB^*))$ is the correlation factor. If unitarity holds, all these correlations define relative phases $\Phi(AB^*)=\Phi(A)-\Phi(B)$ between any pair of complex amplitudes $A$ and $B$ since there are no averaged bilinear terms. In contrast, the measured averaged bilinear terms (8.22) and (8.23) cannot, in general, be represented as real parts of bilinear products $AB^*$ of two complex amplitudes $A$ and $B$. We can see that from the fact that the averaged bilinear terms (8.23) do not satisfy, in general, the necessary condition $(Re(AB^*))^2+(Im(AB^*))^2=|A|^2|B|^2$. This difference allows us to test the validity of the Kraus representation.\\

We will focus on three amplitudes $P^0_\tau$, $D^0_\tau$ and $D^{2U}_\tau$. If unitarity holds, their relative phases must satisfy phase conditions
\begin{equation}
-(\Phi(P^0_\tau)-\Phi(D^0_\tau))+(\Phi(P^0_\tau)-\Phi(D^{2D}_\tau))+
(\Phi(D^{2D}_\tau)-\Phi(D^0_\tau))=0
\end{equation}
or, equivalently, cosine conditions for the measured cosines
\begin{equation}
\cos(\Phi(P^0_\tau D^{0*}_\tau))^2+\cos(\Phi(P^0_\tau D^{2U*}_\tau))^2+
\cos(\Phi(D^{2U}_\tau D^{0*}_\tau))^2
\end{equation}
\[
-2\cos(\Phi(P^0_\tau D^{0*}_\tau)) \cos(\Phi(P^0_\tau D^{2U*}_\tau))
\cos(\Phi(D^{2U}_\tau D^{0*}_\tau))=1
\]
If unitarity is violated, the measurement of averaged bilinear terms will lead to a 
phase gap 
\begin{equation}
-\Phi(P^0_\tau D^{0*}_\tau) + \Phi(P^0_\tau D^{2U*}_\tau)+\Phi(D^{2U}_\tau D^{0*}_\tau)
=\Delta_\tau \neq 0
\end{equation}
and to a cosine gap
\begin{equation}
\cos(\Phi(P^0_\tau D^{0*}_\tau))^2+\cos(\Phi(P^0_\tau) D^{2U*}_\tau))^2+
\cos(\Phi(D^{2U}_\tau D^{0*}_\tau))^2
\end{equation}
\[
-2\cos(\Phi(P^0_\tau D^{0*}_\tau)) \cos(\Phi(P^0_\tau D^{2U*}_\tau))
\cos(\Phi(D^{2U}_\tau D^{0*}_\tau))=\Gamma_\tau \neq 1
\]
The phases $\Phi(P^0_\tau D^{0*}_\tau)$, $\Phi(P^0_\tau D^{2U*}_\tau)$ and
$\Phi(D^{2U}_\tau D^{0*}_\tau)$ were measured at large momentum transfers~\cite{rybicki85} and are shown in Figures 1 and 2. The phase gaps $\Delta_\tau$ and cosine gaps $\Gamma_\tau$ are shown in Figure 3. The cosine gaps were calculated only for the mean values of the phases as it is very difficult to estimate errors on $\Gamma_\tau$ due to nonlinearity in cosines. The phase gaps show large deviations from unitarity value 0. The cosine gaps show large deviations from unitarity value 1.\\ 

Both tests are consistent with the  observation that unitarity conditions (4.15) an (4.16) are violated by the CERN data. Together these facts establish the validity of Kraus representation of the observed final states in pion creation processes. They lead us to a new view of these processes as open quantum systems interacting with a quantum environment.

\section{Decoherence free subspace.}

In this Section we introduce important concepts of decoherence subspace and decoherence free subspace of the Hilbert-Schmidt space of density matrix elements. The concept of decoherence free subspace will be cetral to the determination of quantum states of the environment in a sequel to this work~\cite{svec07b}.\\

Experiments on transversely polarized targets with polarization $\vec {P}=(0,P_y,0)$ measure angular distributions $I^0(\Omega,P_y)$. In terms of (2.15) and (2.40) they have a form
\begin{equation}
I^0(\Omega, P_y)=I^0_u(\Omega)+P_yI^0_y(\Omega)=
\end{equation}
\[
\sum \limits_{J \leq J'} \sum \limits _{\lambda \geq 0} \sum \limits_{\lambda'} 
\xi_{JJ'} \xi_\lambda \bigl ((Re R^0_u)^{JJ}_{\lambda \lambda'}+P_y(Re R^0_y)^{JJ}_{\lambda \lambda'} \bigr ) Re(Y^J_\lambda(\Omega)Y^{J*}_{\lambda'}(\Omega))
\]
Maximum likelihood fits to $I^0(\Omega,P_y)$ with $J\leq J_{max}(m)$ determine density matrix elements $(Re R^0_u)^{JJ}_{\lambda \lambda'}$ and $(Re R^0_y)^{JJ}_{\lambda \lambda'}$, or
equivalently, the density matrix elements for pure initial states $P_y=P_\tau=\pm 1, \tau=u,d$
\begin{equation}
(Re R^0(P_\tau))^{JJ}_{\lambda \lambda'}={1 \over {2}} \bigl ((Re R^0_u)^{JJ}_{\lambda \lambda'}+P_\tau (Re R^0_y)^{JJ}_{\lambda \lambda'} \bigr )
\end{equation}
It follows from Lutz-Rybicki Table 1.d in Appendix A that density matrix elements 
$(Re R^0(P_\tau))^{JJ}_{\lambda \lambda'}$ are related only to amplitudes with the same transversity $\tau =u$ or $\tau=d$. The bilinear terms of amplitudes $P^0_\tau$, $D^0_\tau$ and $D^{2U}_\tau$ discussed in the previous Section are in fact density matrix elements
\begin{equation}
Re(P^0_\tau D^{0*}_\tau)=(Re R^0(P_\tau))^{12}_{00}={1 \over {2}} \sum \limits_{\ell=1}^4
p_{\ell \ell}\bigl ((Re R^0_u(\ell,\ell))^{12}_{00}+P_\tau (Re R^0_y(\ell,\ell))^{12}_{00} \bigr )
\end{equation}
\[
Re(P^0_\tau D^{2U*}_\tau)=(Re R^0(P_\tau))^{12}_{02}={1 \over {2}} \sum \limits_{\ell=1}^4
p_{\ell \ell}\bigl ((Re R^0_u(\ell,\ell))^{12}_{02}+P_\tau (Re R^0_y(\ell,\ell))^{12}_{02} \bigr )
\]
\[
Re(D^{2U}_\tau D^{0*}_\tau)=(Re R^0(P_\tau))^{22}_{20}={1 \over {2}} \sum \limits_{\ell=1}^4
p_{\ell \ell}\bigl ((Re R^0_u(\ell,\ell))^{22}_{20}+P_\tau (Re R^0_y(\ell,\ell))^{22}_{20} \bigr )
\]
where we made use of relations (8.21) to express the measured elements in terms of the co-evolution elements $Re R^0_u(\ell,\ell)$ and $Re R^0_y(\ell,\ell)$. Comparing with (8.20), we can express the co-evolution elements in terms of co-evolution amplitudes
\begin{equation}
(Re R^0_u(\ell,\ell))^{12}_{00}+P_\tau (Re R^0_y(\ell,\ell))^{12}_{00}=Re(P^0_\tau (\ell) D^{0*}_\tau(\ell))
\end{equation}
\[
(Re R^0_u(\ell,\ell))^{12}_{02}+P_\tau (Re R^0_y(\ell,\ell))^{12}_{02}=Re(P^0_\tau (\ell) D^{2U*}_\tau(\ell))
\]
\[
(Re R^0_u(\ell,\ell))^{22}_{20}+P_\tau (Re R^0_y(\ell,\ell))^{22}_{20}=Re(D^{2U}_\tau (\ell) D^{0*}_\tau (\ell))
\]

When the bilinear terms $(Re A^J_\lambda(\ell)B^{J'*}_{\lambda'}(\ell))$ for a pair of co-evolution amplitudes $A^J_\lambda(\ell)$ and $B^{J'}_{\lambda'}(\ell)$ are different for different interacting degrees of freedom of the environment $\ell$, so will be the corresponding co-evolution density matrix elements $(Re R^0_u(\ell,\ell))^{JJ'}_{\lambda \lambda'}+P_\tau (Re R^0_y(\ell,\ell))^{JJ'}_{\lambda \lambda'}$. As the result, the averaged bilinear term $Re(A^J_{\lambda, \tau} B^{J'*}_{\lambda, \tau})$ does not represent a bilinear product of two complex functions and the correlation angle $\Phi(A^J_{\lambda,\tau}B^{J'*}_{\lambda, \tau})$ does not represent a relative phase. We say the corresponding density matrix elements belong to decoherence subspace of the Hilbert-Schmidt space of density matrices.\\

Consider the case of co-evolution amplitudes $P^0_\tau(\ell)$, $D^0_\tau(\ell)$ and $D^{2U}_\tau(\ell)$. They are complex functions and their relative phases must satisfy the phase condition (9.1) and cosine condition (9.2). These conditions are violated by the the measured correlation phases from which we conclude that some of the measured density matrix elements in (10.3) are environment dependent and thus belong to the decoherence subspace.\\

Decoherence free subspace is a subspace of density matrix elements that do not depend on the interaction with the environment. A density matrix element
\begin{equation}
(Re R^0(P_\tau))^{JJ}_{\lambda \lambda'}=Re(A^J_{\lambda, \tau}B^{J'*}_{\lambda',\tau})=
\sum \limits_{\ell=1}^4 p_{\ell \ell} Re (A^J_{\lambda, \tau}(\ell)B^{J'*}_{\lambda',\tau}(\ell))
\end{equation}
will be environment independent only when the bilinear terms 
\begin{equation}
Re(A^J_{\lambda,\tau}(\ell)B^{J'*}_{\lambda',\tau}(\ell))=
|A^J_{\lambda,\tau}(\ell)||B^{J'*}_{\lambda',\tau}(\ell)|
\cos\bigl (\Phi(A^J_{\lambda,\tau}(\ell))-\Phi(B^{J'}_{\lambda',\tau}(\ell))\bigr )
\end{equation}
of its co-evolution amplitudes are all equal and thus do not depend on the environment degrees of freedom $\ell$. In such a case (10.5) takes the form
\begin{equation}
(Re R^0(P_\tau))^{JJ}_{\lambda \lambda'}=Re(A^J_{\lambda, \tau}B^{J'*}_{\lambda',\tau})=
Re(A^J_\lambda(\ell)B^{J'*}_{\lambda'}(\ell))
\end{equation}
where we used $\sum \limits_{\ell=1}^4 p_{\ell \ell}=1$. The relation (10.7) holds for all $\ell$. Note that the moduli and relative phases in (10.6) will, in general, still depend on $\ell$. The measured bilinear term $Re(A^J_{\lambda, \tau}B^{J'*}_{\lambda',\tau})$ now has the form of a bilinear product of two complex functions with a well defined relative phase, albeit not necessarily a unique one.\\

Of central importance is the decoherence free subspace formed by the $S$- and $P$-wave subspace of reduced density matrix $\rho_f^0(\Omega,\vec{P})$~\cite{svec07b}. In this case the trio of amplitudes $S_\tau, P_\tau^0, P_\tau^U$ satisfies the phase and cosine conditions (9.1) and (9.2) which renders the system of equations for moduli and phases analytically solvable. In Ref.~\cite{svec07b} we show that the quantum numbers labeling the solutions can be identified with the quantum states of the environment.\\

\begin{figure}
\includegraphics[width=12cm,height=10.5cm]{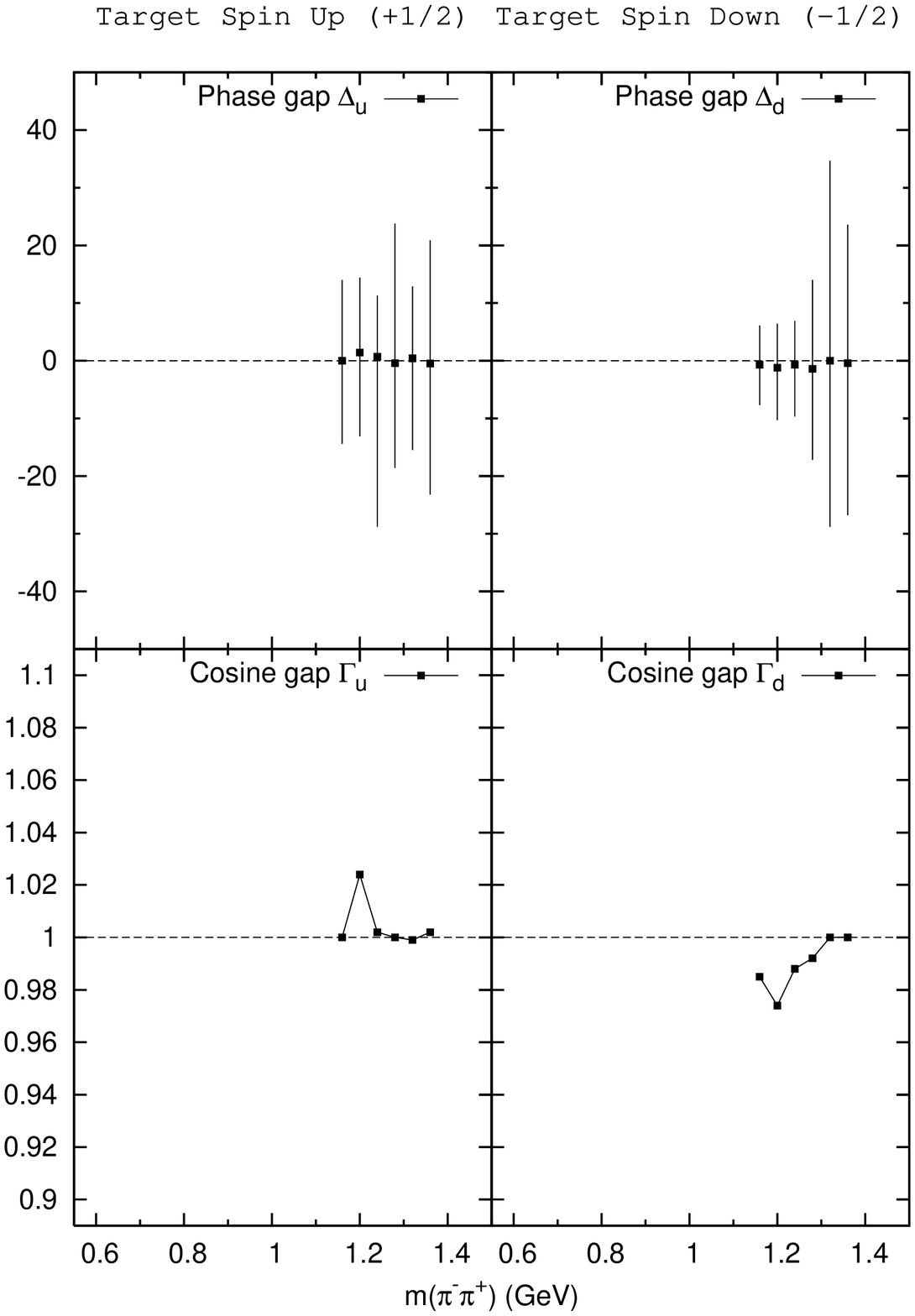}
\caption{Phase gap $\Delta_\tau$ and cosine gap $\Gamma_\tau$ for aplitudes $P^0_\tau$, $D^0_\tau$ and $S_\tau$ at small momentum transfers $0.01 \leq |t| \leq 0.2$ (GeV/c$)^2$. Data from Becker {\sl et al.} ~\cite{becker79b}.}
\label{Figure 4.}
\end{figure}

In general, the $D$-wave subspace of reduced density matrix elements $(\rho^0_k)^{JJ'}_{\lambda \lambda'}$ involving bilinear terms $SD^*$, $PD^*$ and $DD^*$ is not a decoherence free subspace. Elsewhere we shall show that the complete system of $SD^*$, $PD^*$ and $DD^*$ matrix elements measured on transversely polarized target is solvable and yields a unique solution for the $D$-wave co-evolution amplitudes. However, in certain kinematical regions some of these density matrix elements can vanish or be negligible. Such is the case of the CERN analysis~\cite{becker79b}. In the kinematic region for dipion mass 1100-1400 MeV and small momentun transfers $t$ the measurements indicate vanishing of the density matrix elements with dipion helicities $\lambda =\pm2$. In this case the solvability of the complete $D$-wave system requires that the co-evolution amplitudes are equal to their average values independent of $\ell$. At the same time the $P$-wave amplitudes are small and also nearly equal to their average values. As a consequence the same must be true for the $S$-wave amplitudes, the differences between them being three times the diference between $P$-wave amplitudes~\cite{svec07b}. It is now possible for a trio of helicity zero amplitudes $S$, $P^0$ and $D^0$ to satisfy approximately the phase and cosine conditions (9.1) and (9.2). Hence we can write
\begin{equation}
(Re R^0(P_\tau))^{12}_{00}=Re(P^0_\tau D^{0*}_\tau)\sim
|P^0_\tau|| D^{0*}_\tau|\cos(\Phi(P^0_\tau D^{0*}_\tau))
\end{equation}
\[
(Re R^0(P_\tau))^{01}_{00}=Re(S_\tau P^{0*}_\tau)\sim
|S_\tau|| P^{0*}_\tau|\cos(\Phi(S_\tau P^{0*}_\tau))
\]
\[
(Re R^0(P_\tau))^{02}_{00}=Re(S_\tau D^{0*}_\tau)\sim
|S_\tau|| D^{0*}_\tau|\cos(\Phi(S_\tau D^{0*}_\tau))
\]
The correlation phases were measured in amplitude analysis of data at small momentum transfers $0.01 \leq |t| \leq 0.20$ (GeV/c$)$ and presented at dipion masses above 1180 MeV
in Ref.~\cite{becker79b}. The phase gaps and cosine gaps calculated for these phases are shown in Figure 4. The results are very close to 0 and 1, respectively, for both transversities and indicate that the corresponding density matrix elements in (10.8) all belong to approximate decoherence free subspace.\\

Measurements of the correlation phases determine which density matrix elements belong to decoherence or decoherence free subspaces. Both spaces provide information about the nature of the interaction of pion creation processes with the environment.

\section{Quantum states of the environment.}

The obvious question now is what are the quantum states $|e_l>$ of the environment and how do they ensure the diagonal form of co-evolution transversity amplitudes 
$A^{J\eta}_{\lambda, \tau}(\ell m)=A^{J\eta}_{\lambda, \tau}(\ell)\delta_{\ell m}$. In a sequel paper~\cite{svec07b} we report on high resolution amplitude analyses of CERN measurements of $\pi^- p \to \pi^- \pi^+ n$, $\pi^+ n \to \pi^+ \pi^- p$ and $K^+ n \to K^+ \pi^- p$ on transversely polarized targets below dimeson masses $\sim 1000$ MeV where $S$- and $P$-waves dominate. The measured density matrix elements do not form a solvable set of equations for the moduli and cosines of correlation phases. When the data are supplemented by the assumption that the measured phases satisfy the phase relations (9.1), the resulting cosine relations (9.2) provide the missing equations and render the system solvable analytically. This amouts to an assumption that the density matrix elements describing the $S$- and $P$-waves subsystem form a part of the decoherence free subspace.\\

In all processes we find two distinct physical solutions for $S$- and $P$-wave transversity amplitudes $A_u(i)$ and $A_d(j)$, $i,j =1,2$, $A=S,P^0,P^U,P^N$  leading to 4 distinct final states $\rho_f(\Omega, \vec {P}, ij)$. We put forward a hypothesis that all four solutions $\rho_f(\Omega, \vec {P},ij)$ are physical states and that the measured final state $\rho_f(\Omega, \vec {P})$ is a mixed state of these solutions 
\begin{equation}
\rho_f(\Omega, \vec {P})=\sum \limits_{i,j=1}^2 
p_{ij} \rho_f(\Omega, \vec {P},ij)
\end{equation}
where the probabilities $\sum \limits_{i,j=1}^2 p_{ij}=1$.\\

In our next step we associate with the two solutions for transversity amplitudes $A_u(i)$ and $A_d(j)$, $i,j =1,2$ two qubit states $|i>$ and $|j>$, respectively. Then the hypothesis (11.1) allows us to identify the four interacting degrees of freedom of the environment $|e_\ell>$ allowed by the condition $\dim H(E) \leq \dim H_i(S)\dim H_f(S)=(2s_p+1)(2s_n+1)=4$ with the four two-qubit states $|e_\ell> \equiv |i>|j>$. Since the transversity amplitudes can possess only one solution at a time, we get the diagonal form (8.20) for co-evolution amplitudes $A^{J\eta}_{\lambda, \tau}(\ell m)$ for any $J$ and $\lambda$
\begin{equation}
A^{J\eta}_{\lambda, \tau}(\ell m)=A^{J\eta}_{\lambda, \tau}(\ell \ell)\delta_{\ell m}=
A^{J\eta}_{\lambda, \tau}(ij,ij) \delta_{ij,i'j'} \equiv
A^{J\eta}_{\lambda, \tau}(ij) \delta_{ij,i'j'} 
\end{equation}
where
\begin{equation}
A^{J\eta}_{\lambda, u}(ij)=
<J\lambda \eta,\tau_n|<ij|U|ij>|0 u>=A^J_{\lambda, u}(i)
\end{equation}
\[
A^{J\eta}_{\lambda, d}(ij)=
<J\lambda \eta,\tau_n|<ij|U|ij>|0 d>=A^J_{\lambda, d}(j)
\]
Next we identify the probabilities $p_{ij}$ in (11.1) with the diagonal terms $p_{mm}\equiv p_{ij,ij}$ in (8.2). Then the equations (8.22) for the moduli have a modified form
\begin{equation}
|A^{J\eta}_{\lambda, \tau}|^2=\sum \limits_{i,j=1}^2 p_{ij}
|A^{J\eta}_{\lambda, \tau}(ij)|^2
\end{equation}
The equations (8.23) for the measured bilinear terms of transversity amplitudes read
\begin{equation}
Re(A^{J\eta}_{\lambda, \tau}B^{J'\eta'*}_{\lambda', \tau'})\equiv
|A^{J\eta}_{\lambda, \tau}||B^{J'\eta'*}_{\lambda',\tau'}|
\cos(\Phi(A^{J\eta}_{\lambda,\tau}B^{J'\eta'*}_{\lambda', \tau'}))=
\sum \limits_{i,j=1}^2 p_{ij}
Re(A^{J\eta}_{\lambda, \tau}(ij)B^{J'\eta'*}_{\lambda', \tau'}(ij))
\end{equation}
\[
Im(A^{J\eta}_{\lambda, \tau}B^{J'\eta'*}_{\lambda', \tau'})\equiv
|A^{J\eta}_{\lambda, \tau}||B^{J'\eta'*}_{\lambda',\tau'}|
\sin(\Psi(A^{J\eta}_{\lambda,\tau}B^{J'\eta'*}_{\lambda', \tau'}))=
\sum \limits_{i,j=1}^2 p_{ij}
Im(A^{J\eta}_{\lambda, \tau}(ij)B^{J'\eta'*}_{\lambda', \tau'}(ij))
\]
while the equation (8.19) for the measured density matrix elements takes the form
\begin{equation}
(R^n_k)^{JJ'}_{\lambda \lambda'}=\sum \limits_{i,j=1}^2 p_{ij}
(R^n_k (ij,ij))^{JJ'}_{\lambda \lambda'}
\end{equation}
In a sequel paper~\cite{svec07c} we will show that the hypothesis (11.1) is testable experimentally and that the probabilities $p_{ij}$ can be measured.

\section{Time irreversibility and violation of $CPT$ symmetry\\ in pion creation processes.}

The interacting hadrons are matter particles that carry energy-momentum and spin, and so do the two-pion states. The quantum states of the environment
\begin{equation}
\rho_i(E)=\sum \limits_{ij,i'j'}p_{ij,i'j'}|i>|j><i'|<j'|
\end{equation}
can be thought of as particles carrying quantum entanglement. The unitary co-evolution $U$ of the pion creation process with the environment can be interpreted as the scattering of the initial hadron state $\rho_i(S)=\rho_i(\pi^-p,\vec {P})$ with the quantum state $\rho_i(E)$ of the environment 
\begin{equation}
\rho_i(S) \otimes \rho_i(E) \to^U \rho_f(S,E)
\end{equation}
where $\rho_f(S,E)$ is the entangled state of the final hadron system $\pi^-\pi^+n$ with the environment $E$. When observers perform measurements of the final state $\rho_f(S,E)$ they make use of the amplitude analysis as an integral part of their measurement and conclude that the observed state $\rho_f(\pi^-\pi^+n, \vec {P})$ is a mixed state even for pure initial states $\pi^-p$. This observation informs them about the existence of the quantum environment. The observers cannot use their classiclal instruments to observe the environment. It is the pion creation processes that act as non-classical instruments which allow observers to access information about the nature of the environment and its interactions with hadron processes.\\

As we show in a sequel paper~\cite{svec07b}, the observed state is a mixed state of solutions produced by amplitude analysis of data on polarized targets
\begin{equation}
\rho_f(\pi^-\pi^+n, \vec {P})=\sum \limits_{i,j=1}^2 p_{ij}\rho_f(\pi^-\pi^+n, \vec {P},ij)
\end{equation}
The interaction with the environment leads to a "level splitting" of the probability amplitudes into solutions $A^{J\eta}_{\lambda,u}(i)$ and 
$A^{J\eta}_{\lambda,d}(j)$, $i,j=1,2$ . Such splitting cannot be prepared by the observers in the initial state of the time reversed process $\pi^-\pi^+n \to \pi^- p$ and the pion creation process $\pi^-p \to \pi^- \pi^+ n$ is thus time-irreversible. Also, it is not possible for observers to prepare the entangled state $\rho_f(S,E)$ as an initial state that would undo the entanglement of the pion creation process $S$ and the environment $E$ and recreate the initial separable state $\rho_i(S) \otimes \rho_i(E)$.\\

According to $CPT$ Theorem, in any local and Lorentz invariant field theory in Minkovski spacetime the vacuum expectation values of time-ordered field operators are $CPT$ invariant~\cite{streater00}. Assuming in addition the unitary $S$-matrix (1.2), the $CPT$ invariance is extended to the observable $S$-matrix amplitudes~\cite{luders57}. The $CPT$ invariance then means that the apmlitude $A$ which describes a process $a+b \to c+d$ also describes a $CPT$ conjugate process
$\bar {c} + \bar {d} \to \bar {a} + \bar {b}$. This is possible only when masses, widths, spins and isospins of particles and antiparticles are equal.\\

Interactions that are invariant under $CPT$ symmetry lead to observables that are invariant under the $CPT$ as well. Such is the case for interactions that give rise to hadron masses, widths, spins and isospins. Interactions that violate $CPT$ symmetry lead to observables that violate $CPT$ symmetry. Interactions (12.2) of pion creation processes with environment violate $CPT$ symmetry because of the non-local character of the quantum states $\rho_i(E)$ and because these states do not have well defined antiparticle states. Consequently, the amplitudes $A^{J\eta}_{\lambda, \tau}(ij)$ in (11.3) for the process
\begin{equation}
|\pi^-p> + |i>|j> \to |\pi^-\pi^+n> +|i>|j>
\end{equation}
do not describe the process 
\begin{equation}
|\pi^+ \pi^- \bar {n}> +|i>|j> \to |\pi^+ \bar {p}> + |i>|j> 
\end{equation}
Since the co-evolution operator $U$ is unitary, the processes (12.2) must be logically time-reversible even though they violate $CPT$ symmetry and are time-irreversible from the point of view of observers of the reduced states $\rho_f(S)$ who cannot prepare the unknown states $\rho_f(S,E)$.\\ 

It is important to recognize that $CPT$ violating interactions with environment need not contradict or affect other $CPT$ invariant interactions and their observables in the same process. Thus we may expect the masses and widths of $\pi^-$ and $\pi^+$ to be the same. The $CPT$ violating interactions will manifest themselves in other observable aspects of the pion creation processes. In a series of sequel papers on high resolution amplitude analyses we will show that the interaction with environment manifests itself in mixing of scalar and vector resonances below $\sim 1000$ MeV in both $\pi N \to \pi^- \pi^+ N$~\cite{svec07b} and $K^+ n \to K^+ \pi^- p$ processes, and in violation of the Generalized Bose-Einstein symmetry in $\pi^- p \to \pi^- \pi^+ n$~\cite{svec07b}. In previous low resolution analyses of $\pi N \to \pi^- \pi^+ N$ the presence of $\rho^0(770)$ resonance in the $S$-wave amplitudes was misinterpreted as evidence for a scalar resonance 
$\sigma (770)$~\cite{svec92b,svec96,svec97a,svec02a,alekseev99}.\\

There is no energy-momentum of the quantum state $\rho_i(E)$ of the environment involved in the interaction with pion creation processes. The interacting hadrons conserve their energy-momentum and there is no exchange of energy-momentum with the environment, in agreement with the original proposal by Hawking for particle processes interacting with quantum fluctuations of space-time metric~\cite{hawking82,hawking84}. Instead, the interaction with environment is a non-dissipative (dephasing) process and involves change of quantum entanglement of quantum states of the environment and the produced hadrons. The entanglement changing interaction modifies the entanglement content of the quantum state of the environment to $\rho_f(E)=Tr_{S}(\rho_f(S,E))$ and stores quantum information in dimeson states observable as mixing of scalar-vector resonances~\cite{svec07b} and as violation of Generalized Bose-Einstein symmetry~\cite{svec07b}. This last observation seems to confirm the change of entanglement of particle-antiparticle pairs called "$\omega$-effect" and recently predicted by Bernab\'{e}u, Mavromatos and Sarkar to arise in $CPT$ violating interactions of maximally entangled particle-antiparticle pairs with space-time foam~\cite{bernabeu06a,bernabeu06b}. 

\section{Conclusions.}

It is the task of spin physics to use known initial spin states $\rho_i$ to measure and to investigate the final spin states $\rho_f$ to learn about new aspects of the dynamics of hadron interactions. It is therefore not surprising that the CERN measurements of pion creation processes on polarized targets provide information about the validity of the unitarity assumptions (1.1) and (1.2). The observed violation of the unitarity conditions 
(4.15) and (4.16) means that pure initial states can evolve into mixed final states in these processes. This leads to a new view of pion creation processes as open quantum systems interacting with a quantum environment in a unitary co-evolution. The non-unitary evolution from the initial state to the observed final state is described by Kraus representation that leaves invariant the form of measured angular distributions in terms of environment averaged density matrix elements.\\

The measured process is time-irreversible and violates $CPT$ symmetry. Due to the non-local character of the interaction with the environment and due to the fact that the quantum states $\rho_i(E)$ of the environment do not have well defined antiparticle states, the co-evolution process itself violates $CPT$ symmetry. The interaction with the environment manifests itself in scalar-vector mixing and in violation of Bose-Einstein symmetry.\\

The CERN measurements of pion creation processes on polarized targets opened a window on a entirely new class of phenomena which result from their interactions with a quantum environment. Following Hawking's proposals, these phenomena can be understood as low energy manifestations of quantum gravity. In this view, the observed quantum environment originates in quantum gravity and the pion creation processes act as non-classical instruments that make quantum gravity an observable reality.

\acknowledgements

Krzysztof Rybicki pioneered the measurement and the amplitude analysis of the pion creation processes on polarized targets at CERN. Over many years Krzysztof Rybicki maintained interest in my study of the CERN data and helped me with his advice. Krzysztof died in March 2003. I wish to dedicate this work to the Memory of Krzysztof Rybicki.

\appendix
\section{Lutz-Rybicki tables of density matrix elements.}

In their 1978 paper~\cite{lutz78}, Lutz and Rybicki tabulated expressions for all 16 kinds of angular density matrix elements $(R^j_k)^{JJ"}_{\lambda \lambda'}$, $j=0,1,2,3$, $k=u,x,y,z$ in terms of helicity amplitudes, helicity amplitudes with definite naturality, nucleon transversity amplitudes and nucleon transversity amplitudes with definite naturality. Their work has been used in all amplitude analyses of CERN data on $\pi N \to \pi^- \pi^+ N$ and $K^+ n \to K^+ \pi^- p$ on polarized 
targets~\cite{becker79a, chabaud83,rybicki85,svec92a,svec92c,svec96,svec97a} but was not published. Since their paper is no longer available, we reproduce their tables for helicity amplitudes and nucleom transversity amplitudes with definite naturality, which are used in this work, in Figures 5 and 6, respectively. In our notation their nucleon helicity nonflip and flip amplitudes read
\begin{equation}
N^j_m=H^J_{\lambda +,0+}, \quad F^j_m=H^J_{\lambda+,0-}
\end{equation}
The relations for nucleon transversity amplitudes with definite naturality read
\begin{equation}
{}^Ug^j_m=U^J_\lambda=U^J_{\lambda,u},\quad {}^Uh^j_m =\overline {U}^J_\lambda=U^J_{\lambda,d}
\end{equation}
\[
{}^Ng^j_m=\overline {N}^J_\lambda=N^J_{\lambda,d},\quad {}^Nh^j_m=N^J_\lambda =N^J_{\lambda,u}
\]
In their notation, the parity relations (2.52) for transversity amplitudes with definite naturality are
\begin{equation}
{}^Ug^j_{-m}=+(-1)^m {}^Ug^j_m, \quad {}^Uh^j_{-m}=+(-1)^m {}^Uh^j_m
\end{equation}
\[
{}^Ng^j_{-m}=-(-1)^m {}^Ng^j_m, \quad {}^Nh^j_{-m}=-(-1)^m {}^Nh^j_m
\]
They use lower case $jj'$ and $mm'$ while we use $JJ'$ and $\lambda \lambda'$ with the same meaning.\\

Nucleon helicity amplitudes with definite $t$-channel naturality defined in (2.44) and (2.45) are better suited for certain theoretical studies. Their characteristic feature is pion exchange dominance of unnatural helicity flip amplitudes at small momentum transfers $t$. In the companion paper~\cite{svec07b} we show how conversion of transversity amplitudes into helicity amplitudes with definite naturality determines the relative phase of transversity amplitudes with opposite nucleon transversity and thus recoil nucleon polarization. In Figure 7 we reproduce Lutz-Rybicki Table 1b for nucleon helicity amplitudes with definite $t$-channel naturality. In our notation their amplitudes read
\begin{equation}
{}^Un^j_m=U^J_{\lambda+,0+}, \qquad {}^Nn^j_m=N^J_{\lambda+,0+}
\end{equation}
\[
{}^Uf^j_m=U^J_{\lambda+,0-}, \qquad {}^Nf^j_m=N^J_{\lambda+,0-}
\]
Their transversity amplitudes (A2) are related to their helicity amplitudes (A4) by relations 
\begin{equation}
{}^U g^j_m={1 \over {\sqrt{2}}}({}^U n^j_m+i{}^U f^j_m), \quad
{}^U h^j_m={1 \over {\sqrt{2}}}({}^U n^j_m-i{}^U f^j_m)
\end{equation}
\[
{}^N g^j_m={1 \over {\sqrt{2}}}({}^N n^j_m-i{}^N f^j_m), \quad
{}^N h^j_m={1 \over {\sqrt{2}}}({}^N n^j_m+i{}^N f^j_m)
\]
\[
{}^N g^j_0={}^N h^j_0=0
\]
These are the same relations as (2.49) and (2.50). To achieve this conformity was the reason why we omitted the factor $i$ in (2.47).

\begin{figure}
\includegraphics[width=16cm,height=21.5cm]{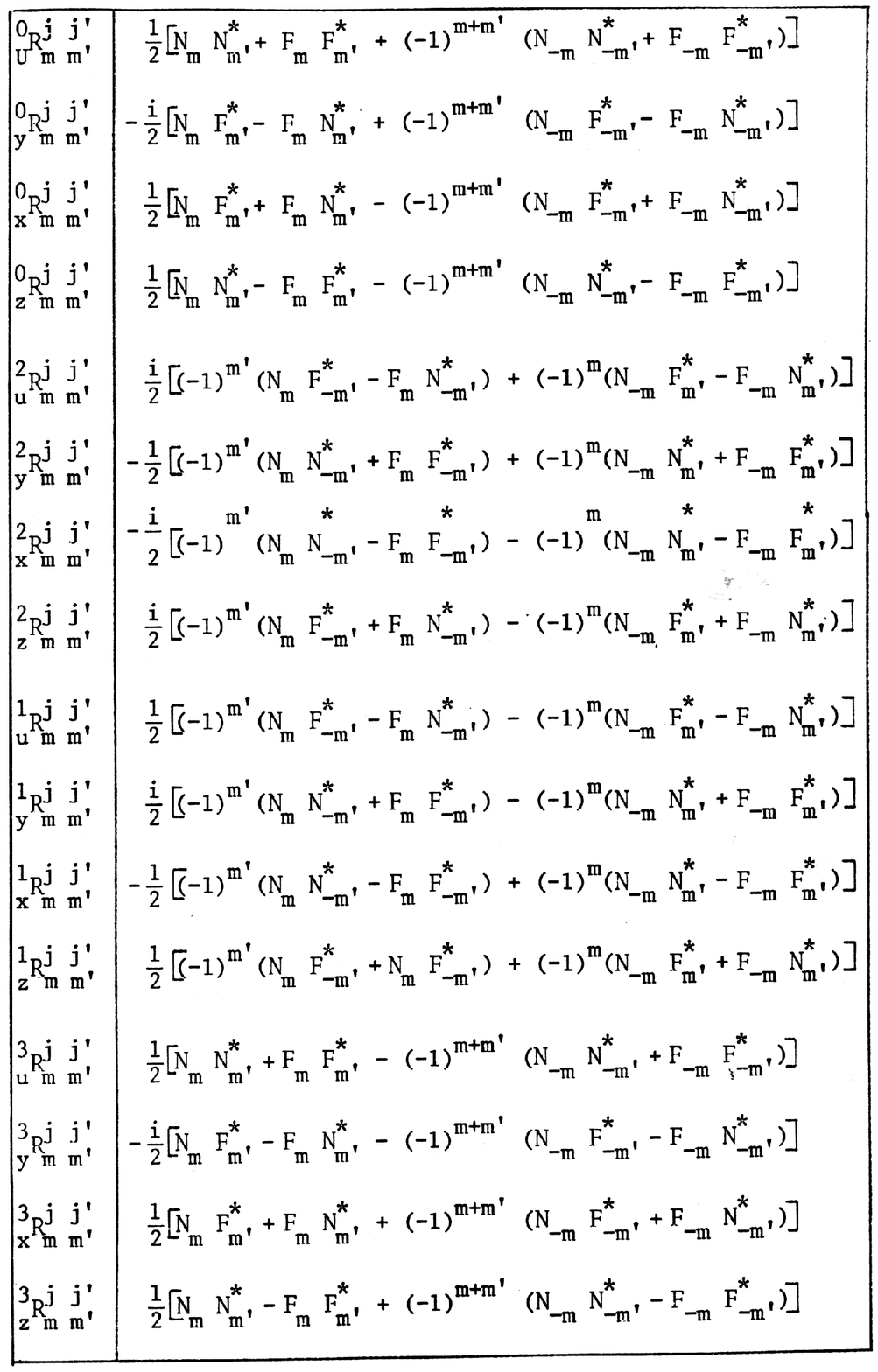}
\caption{Angular density matrix elements expressed in terms of helicity amplitudes. The spin indices $jj'$ which always go with $mm'$ have been omitted in the amplitudes. Reproduced Table 1a of Ref.~\cite{lutz78}.}
\label{Table IA}
\end{figure}

\begin{figure}
\includegraphics[width=16cm,height=21.5cm]{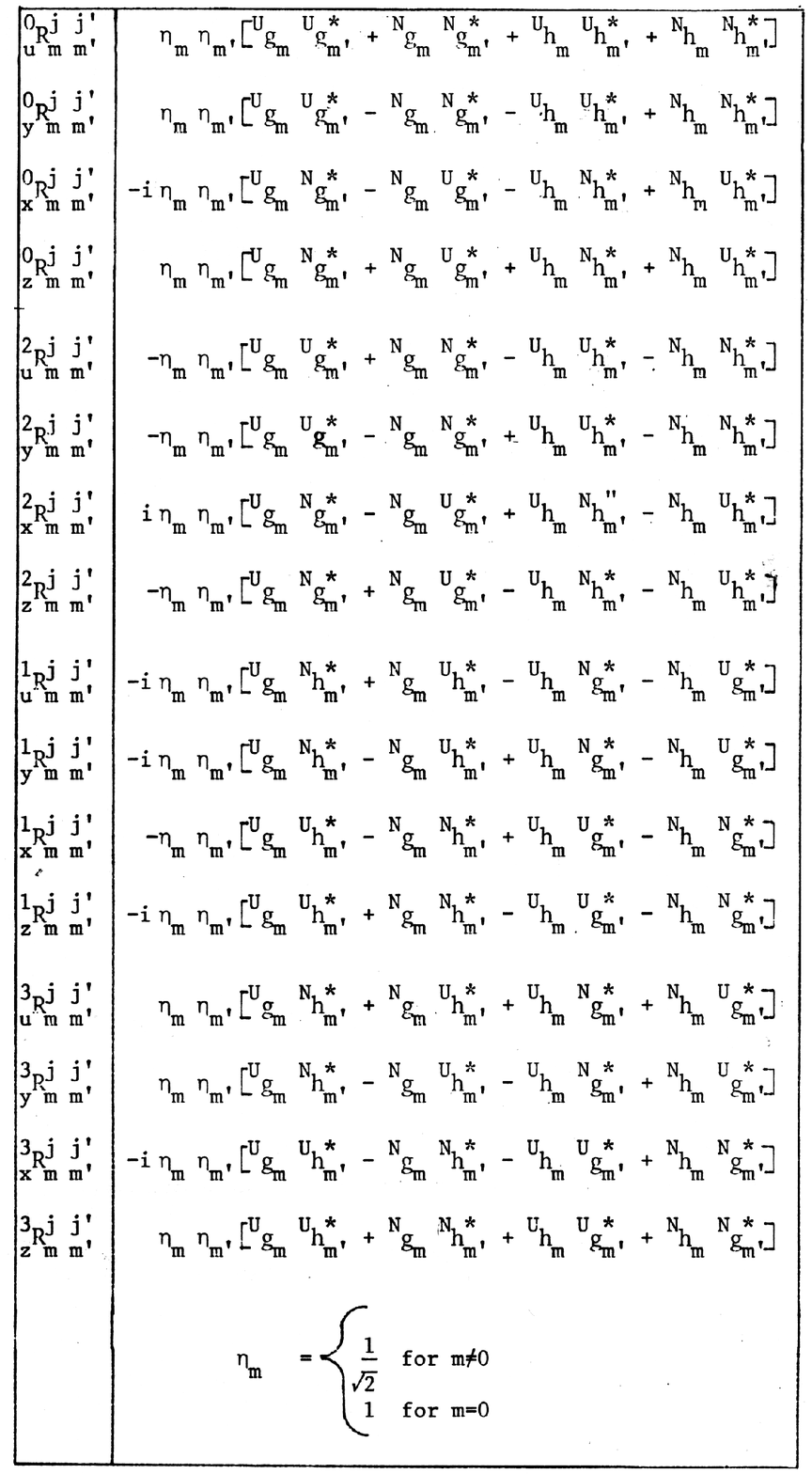}
\caption{Angular density matrix elements expressed in terms of nucleon transversity amplitudes with definite $t$-channel naturality. The spin indices $jj'$ which always go with $mm'$ have been omitted in the amplitudes. Reproduced Table 1d of Ref.~\cite{lutz78}.}
\label{Table IIA}
\end{figure}

\begin{figure}
\includegraphics[width=16cm,height=21.5cm]{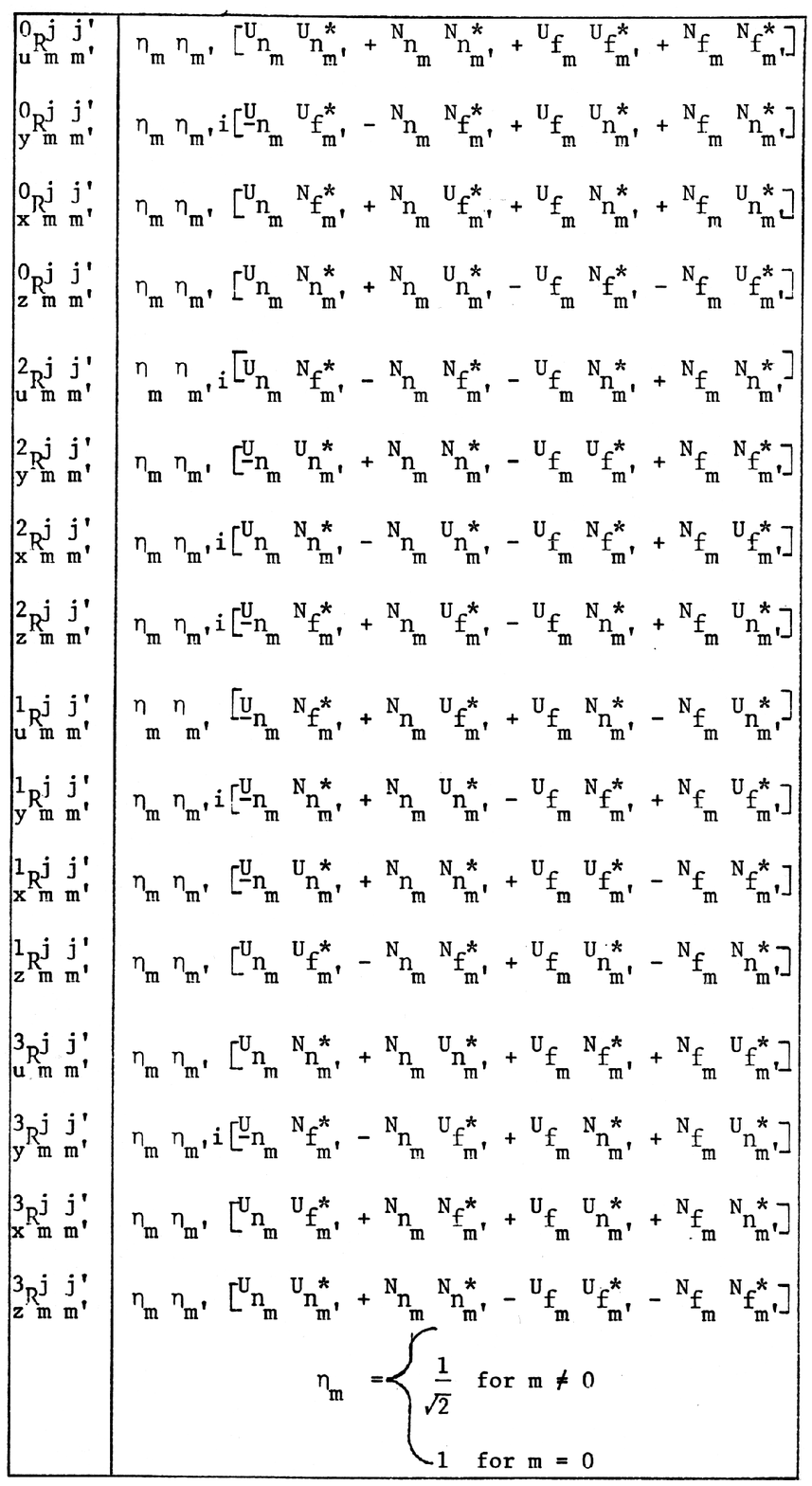}
\caption{Angular density matrix elements expressed in terms of nucleon helicity amplitudes with definite $t$-channel naturality. The spin indices $jj'$ which always go with $mm'$ have been omitted in the amplitudes. Reproduced Table 1b of Ref.~\cite{lutz78}.}
\label{Table IIIA}
\end{figure}

\end{document}